\documentclass[aip,graphicx]{revtex4-1}

\usepackage{mathtools}
\usepackage{mathrsfs}
\usepackage{esint}
\usepackage{xcolor}
\usepackage{tikz}
\usepackage[shortlabels]{enumitem}

\usepackage[margin=3.4cm]{geometry}

\usepackage{amsmath}
\usepackage{amsfonts}
\usepackage{amsthm}
\usepackage{amssymb}
\usepackage{bbm}

\usepackage{setspace}

\usepackage{titlesec}
\titleformat{\section}{\Large\bfseries}{\thesection}{1.5ex}{}
\titleformat{\subsection}{\large\bfseries}{\thesubsection}{1.5ex}{}

\usepackage{graphicx}
\usepackage{booktabs}
\usepackage[suffix=]{epstopdf}
\usepackage[font=footnotesize,labelfont=bf]{caption}
\usepackage[colorlinks]{hyperref}
\hypersetup{citecolor=blue,urlcolor=blue,linkcolor=blue}
\newtheorem{lemma}{Lemma}

\newtheorem*{conjecture*}{Conjecture}
\theoremstyle{definition}

\theoremstyle{remark}

\usepackage{cleveref}
\crefname{equation}{}{}
\crefname{enumi}{}{}
\crefname{theorem}{Theorem}{Theorems}
\crefname{corollary}{Corollary}{Corollaries}
\crefname{example}{Example}{Examples}
\crefname{lemma}{Lemma}{Lemmas}
\crefname{proposition}{Proposition}{Propositions}
\crefname{figure}{Figure}{Figures}
\crefname{table}{Table}{Tables}
\crefformat{section}{\S#2#1#3}
\crefmultiformat{section}{\S#2#1#3}{ and~\S#2#1#3}{, \S#2#1#3}{, and~\S#2#1#3}
\crefformat{subsection}{\S#2#1#3}
\crefmultiformat{subsection}{\S#2#1#3}{ and~\S#2#1#3}{, \S#2#1#3}{, and~\S#2#1#3}
\crefname{appendix}{appendix}{appendices}
\Crefname{equation}{}{}
\Crefname{enumi}{}{}
\Crefname{theorem}{Theorem}{Theorems}
\Crefname{corollary}{Corollary}{Corollaries}
\Crefname{example}{Example}{Examples}
\Crefname{lemma}{Lemma}{Lemma}
\Crefname{proposition}{Proposition}{Proposition}
\Crefname{figure}{Figure}{Figures}
\Crefname{table}{Table}{Tables}
\Crefformat{section}{Section~#2#1#3}
\Crefmultiformat{section}{Sections~#2#1#3}{ and~#2#1#3}{, #2#1#3}{, and~#2#1#3}
\Crefformat{subsection}{Section~#2#1#3}
\Crefmultiformat{subsection}{Sections~#2#1#3}{ and~#2#1#3}{, #2#1#3}{, and~#2#1#3}
\Crefname{appendix}{Appendix}{Appendices}

\newcommand{\uvec}[1]{\boldsymbol{\hat{\mathbf{#1}}}}
\newcommand\Ra{\mbox{\textit{R}}}  
\renewcommand\Pr{\mbox{\textit{Pr}}}  
\newcommand{\wT}{\mbox{$\smash{\mean{wT}}$}}

\newcommand{\vx}{\boldsymbol{x}}
\newcommand{\vu}{\boldsymbol{u}}

\newcommand{\volav}[1]{\left\langle #1 \right\rangle}

\newcommand{\horav}[1]{\langle #1 \rangle_h}
\newcommand{\timeav}[1]{\overline{#1}}
\newcommand{\mean}[1]{\timeav{\volav{#1}}}
\newcommand{\bR}{\mathbb{R}}
\newcommand{\Tspace}{\mathcal{H}}

\newcommand{\bfield}{\tau} 
\newcommand{\lm}{\lambda} 
\newcommand{\bp}{\beta} 

\newcommand{\subeqref}[2]{\hyperref[#1]{(\ref*{#1}#2)}}

\definecolor{matlabblue}{RGB}{0,113,188}
\definecolor{myblue}{RGB}{0,0,250}
\definecolor{matlabred}{RGB}{216,82,24}
\definecolor{mygrey}{rgb}{0.7,0.7,0.7}
\definecolor{matlabgreen}{rgb}{0.4660,0.6740,0.1880} 
\definecolor{lightblue}{RGB}{0,180,200}
\definecolor{lightred}{RGB}{250,100,40}

\definecolor{white}{rgb}{0 0 0}
\definecolor{matlabyellow}{rgb}{0.93,0.69,0.13} 
\definecolor{colorbar1}{rgb}{1.000000,0.909091,0.000000}
\definecolor{colorbar2}{rgb}{1.000000,0.818182,0.000000}
\definecolor{colorbar3}{rgb}{1.000000,0.727273,0.000000}
\definecolor{colorbar4}{rgb}{1.000000,0.636364,0.000000}
\definecolor{colorbar5}{rgb}{1.000000,0.545455,0.000000}
\definecolor{colorbar6}{rgb}{1.000000,0.454545,0.000000}
\definecolor{colorbar7}{rgb}{1.000000,0.363636,0.000000}
\definecolor{colorbar8}{rgb}{1.000000,0.272727,0.000000}
\definecolor{colorbar9}{rgb}{1.000000,0.181818,0.000000}
\definecolor{colorbar10}{rgb}{1.000000,0.090909,0.000000}
\definecolor{colorbar11}{rgb}{1.000000,0.000000,0.000000}
\definecolor{colorbar12}{rgb}{0.909091,0.000000,0.000000}
\definecolor{colorbar13}{rgb}{0.818182,0.000000,0.000000}
\definecolor{colorbar14}{rgb}{0.727273,0.000000,0.000000}
\definecolor{colorbar15}{rgb}{0.636364,0.000000,0.000000}
\definecolor{colorbar16}{rgb}{0.545455,0.000000,0.000000}
\definecolor{colorbar17}{rgb}{0.454545,0.000000,0.000000}
\definecolor{colorbar18}{rgb}{0.363636,0.000000,0.000000}
\definecolor{colorbar19}{rgb}{0.272727,0.000000,0.000000}
\definecolor{colorbar20}{rgb}{0.181818,0.000000,0.000000}
\definecolor{colorbar21}{rgb}{0.090909,0.000000,0.000000}
\definecolor{grey}{rgb}{0.7,0.7,0.7}

\newcommand\solidrule[1][10pt]{\rule[0.5ex]{#1}{1.5pt}}
\newcommand\dashedrule{\mbox{\solidrule[2pt]\hspace{2pt}\solidrule[2pt]\hspace{2pt}\solidrule[2pt]}}

\newcommand\dottedrule{\mbox{\solidrule[1pt]\hspace{1pt}\solidrule[1pt]\hspace{1pt}\solidrule[1pt]\hspace{1pt}\solidrule[1pt]\hspace{1pt}\solidrule[1pt]\hspace{1pt}}}

\newcommand{\mysquare}[1]{%
	\protect\begin{tikzpicture}%
	\protect\fill [color=#1] (0,0) -- (0.75ex,0) -- (0.75ex,0.75ex) -- (0,0.75ex) -- (0,0);
	\protect\end{tikzpicture}%
}

\newcommand{\mycross}[1]{%
	\protect\begin{tikzpicture}%
	\protect\draw[thick,color=#1] (0,0) -- (1ex,1ex);
	\protect\draw[thick,color=#1] (0,1ex) -- (1ex,0);
	\protect\end{tikzpicture}%
}
\newcommand{\mycirc}[1]{%
	\protect\begin{tikzpicture}%
	\protect\draw[thick,color=#1] (0.5ex,0.5ex) circle (0.5ex);
	\protect\end{tikzpicture}%
}

\draft 

\begin{document}


\title{New bounds for heat transport in internally heated convection at infinite Prandtl number} 


\author{Ali Arslan}
\email{ali.arslan@erdw.ethz.ch}
\affiliation{%
 Institute of Geophysics, ETH Z\"{u}rich, Z\"{u}rich, CH-8092, Switzerland \\
}%
%
\author{Rub\'en E. Rojas}
\affiliation{%
 Institute of Geophysics, ETH Z\"{u}rich, Z\"{u}rich, CH-8092, Switzerland \\
}%

 

%



\date{\today}

\begin{abstract}

We prove new bounds on the heat flux out of the bottom boundary, $\mathcal{F}_B$, for a fluid at infinite Prandtl number, heated internally between isothermal parallel plates under two kinematic boundary conditions. In uniform internally heated convection, the supply of heat equally leaves the domain by conduction when there is no flow. When the heating, quantified by the Rayleigh number, $R$, is sufficiently large, turbulent convection ensues and decreases the heat leaving the domain through the bottom boundary. In the case of no-slip boundary conditions, with the background field method, we rigorously determine that $\mathcal{F}_B \gtrsim R^{-2/3} - R^{-1/2}\log{(1-R^{-2/3})}$ up to a positive constant independent of the Rayleigh and Prandtl numbers. Whereas between stress-free boundaries we prove that $\mathcal{F}_B \gtrsim R^{-40/29} - R^{-35/29}\log{(1-R^{-40/29})}$. We perform a numerical study of the system in two dimensions up to a Rayleigh number of $5\times10^9$ with the spectral solver Dedalus. Our numerical investigations indicate that $\mathcal{F}_B \sim R^{-0.092} $ and $\mathcal{F}_B \sim R^{-0.12}$ for the two kinematic boundary conditions respectively. The gap between the scaling in our simulations and our constructions in the proof indicates that further optimisation could improve the rigorous bounds on $\mathcal{F}_B$.

\end{abstract}


\maketitle 


\section{ Introduction}
\label{sec:intro}

Turbulent convection is ubiquitous in nature, be it atmospheric convection, mixing in lakes or mantle convection within planets; the motion of fluids shapes the physics of the Earth. Studies of turbulent convection use the model introduced by Lord Rayleigh \cite{LordRayleigh1916}, referred to as Rayleigh-B\'enard convection (RBC), where temperature variations are generated by heating a fluid confined between two plates from the lower boundary. However, for turbulent convection in geophysics, heat is generated and removed throughout the domain \cite{sparrow1964,roberts1967convection,tritton_zarraga_1967}. For atmospheres and lakes, this occurs by the absorption of solar radiation \cite{emanuel1994atmospheric,pierrehumbert2010,seager2010}, whereas, for the Mantle, the radioactive decay of isotopes provides a constant supply of heat for convection \cite{schubert2015treatise,schubert2001mantle}.   

By numerical simulations or experiments, emergent quantities of turbulent convection, like the mean vertical heat transport, can be studied. However, experiments and simulations currently cannot reach parameter values of relevance to geophysics. The Prandtl number quantifying the ratio of thermal to viscous diffusion of a fluid varies significantly within planets from $10^{23}$ in the Mantle to $10^{-1} $ in the liquid core \cite{schubert2015treatise}. Similarly, the Rayleigh number, $R$, quantifying the destabilising effect of internal heating to the stabilising effect of diffusion, is estimated to be at least $10^{6}$ in the Mantle and possibly $10^{29}$ in the core \cite{bercovici2011mantle,GUBBINS20013}.

A mathematically rigorous study of turbulent convection can instead give insight into heat transport at all parameter values. In individual papers, Malkus, Howard and Busse (MHB) \cite{Malkus1954,Howard1972,Busse1969}, working with the premise that turbulence maximises heat transport, constructed a variational method to determine the dependence of mean convective heat transport, $\wT$, to $R$ (overlines denote an infinite time and the angled brackets a volume average). Rather than
optimising $\wT$ over solutions to the governing equations, one optimises over the set of incompressible flow fields that only satisfy integral constraints in the form of energy balances. The optima of the MHB approach is still difficult to evaluate. In the 1990s, Doering and Constantin demonstrated that conservative one-sided bounds are possible with their \textit{background method} \cite{Doering1992,Doering1994,Constantin1995a,Doering1996}. The problem reduces from a variational problem over the state variables into one of constructing a background field. 
The background method has enjoyed significant success since its introduction \cite{Fantuzzi2022}.

Recent insights have demonstrated that the background method and alternative bounding methodologies can be systematically formulated within a more general framework for bounding infinite-time averages, called the \textit{auxiliary functional method} \cite{Chernyshenko2022,Chernyshenko2014a}. Using quadratic functionals makes the approach equivalent to the background method. This auxiliary functional method yields sharp bounds for well-posed ordinary and partial differential equations under certain technical conditions \cite{Tobasco2018,Rosa2020}. The advantages of this formulation are that:
(i) it provides a systematic approach to bound any mean quantity of choice, and (ii) it reduces the search for a bound to a convex variational principle. Furthermore, the variational problem can be simplified using symmetries and improved by incorporating additional constraints such as maximum/minimum principles \cite{Fantuzzi2022}.

Building upon previous work on IHC, this paper applies the auxiliary functional method to uniform internally heated convection (IHC) at infinite $Pr$ between isothermal plates. The main results are a bound for no-slip boundaries that improve on the work of Arslan et al.(2023)\cite{Arslan2023} and a novel bound for stress-free boundaries on the change in heat flux out of the domain due to turbulent convection. The bounds are compared with a two-dimensional numerical study of IHC obtained with the spectral solver DEDALUS. Our mathematical results take inspiration from previous applications of the background method to RBC and reveal that, unlike RBC, there remains scope for improvement of the bounds for flows driven by internal heating at infinite $Pr$. 

In RBC, the Nusselt number, $Nu=1+\wT$, defines heat transport enhancement by convection, and the best-known bounds for RBC are optimal within the background method \cite{Wen2015a,Nobili2022}. For arbitrary $Pr$, it was established that $Nu \lesssim Ra^{1/2}$, where $Ra$ is the Rayleigh number in RBC quantifying the destabilising effect of boundary heating to diffusion \cite{Doering1996}. Here, $\lesssim$ and $\gtrsim$ denote bounds that hold up to positive constants independent of $Pr$, $Ra$, the initial conditions or aspect ratio. In the limit of infinite $Pr$, it was proven, that, $Nu \lesssim Ra^{1/3}$ up to logarithms \cite{Doering2006}. The $Ra^{1/2}$ and $Ra^{1/3}$ scaling represent two different phenomenological predictions, known as the classical and ultimate regimes of turbulent heat transport. With alternative techniques to the background method, it has been demonstrated, up to logarithms, that $Nu \lesssim Ra^{1/3}$  provided that $Pr \gtrsim Ra^{1/3}$ and $Nu \lesssim Pr^{-1/2} Ra^{1/2}$ otherwise \cite{choffrut2016upper}. While the variation of the thermal boundary conditions does not alter the results, changing the kinematic boundary conditions does. Between stress-free boundaries, it has been proven, that $Nu \lesssim Ra^{5/12}$, by exploiting additional information in the enstrophy\cite{Whitehead2011prl,wang2013,Whitehead2012}. As such, the ultimate regime does not exist for RBC when $Pr$ is higher than $Ra^{1/3}$ or if the boundaries are stress-free. While bounds on IHC can also provide insight into the ultimate state of turbulent convection, internal heating introduces additional features such that bounds instead give insight into the limits of energy estimates on the underlying PDEs.   

The two emergent quantities of turbulent convection are  $\wT$ and the mean temperature $\mean{T}$, unlike RBC in IHC, the two cannot be \textit{a priori} related\cite{Goluskin2016book}. The bounds for RBC translate over to IHC when bounding $\mean{T}$, which is interpreted as $Nu^{-1}$, albeit this relation is only empirical \cite{Goluskin2016book}. It was established that for arbitrary $Pr$ that $\mean{T}\gtrsim R^{-1/3}$ between no-slip boundaries\cite{Lu2004}, while at $Pr = \infty$, $\mean{T} \gtrsim R^{-1/4}$ up to logarithms\cite{Whitehead2011}. In the case of stress-free boundaries, the background method gives that $\mean{T} \gtrsim R^{-5/17}$ by use of the same approach as in RBC \cite{Whitehead2012}. The proofs on $\mean{T}$ do not translate directly to $\wT$ in IHC. Instead, to obtain bounds on $\wT$ with the background method, one must enforce a minimum principle on the temperature, which states that the temperature within the domain remains above the value prescribed at the boundaries \cite{Arslan2021}. Furthermore, recent work has determined the need for background fields of higher complexity, where the boundary layers are of different widths and can have multiple layers\cite{Arslan2021a,kumar2021ihc,Arslan2023}.

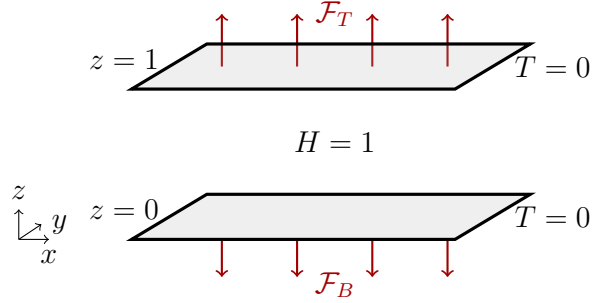
\begin{figure}
    \centering
    \begin{tikzpicture}[every node/.style={scale=0.95}]

    \draw [->,colorbar15,thick] (-0.8,0) -- (-0.8,-0.5);
    \draw [->,colorbar15, thick]  (0.2,0) -- (0.2,-0.5);
    \draw [->,colorbar15, thick] (1.2,0) -- (1.2,-0.5);
    \draw [->,colorbar15, thick ]  (2.2,0) -- (2.2,-0.5);
    \draw [->,colorbar15, thick  ]  (-0.8,2.3) -- (-0.8,3);
    \draw [->,colorbar15 , thick ]  (0.2,2.3) -- (0.2,3);
    \draw [->,colorbar15, thick ]  (1.2,2.3) -- (1.2,3);
    \draw [->,colorbar15, thick ]  (2.2,2.3) -- (2.2,3);
    \draw[black,very thick, fill=mygrey, fill opacity = 0.2] (-2,2) -- (-1,2.6) -- (3.3,2.6) -- (2.3,2) -- cycle;
    \draw[black,very thick, fill=mygrey, fill opacity = 0.2] (-2,0) -- (-1,0.6) -- (3.3,0.6) -- (2.3,0) -- cycle;
    \draw[->] (-3.5,0) -- (-3.1,0) node [anchor=north] {$x$};
    \draw[->] (-3.5,0) -- (-3.2,0.2) node [anchor=west] {$y$};
    \draw[->] (-3.5,0) -- (-3.5,0.4) node [anchor=south] {$z$};
    \node at (-2.1,0.4) {$z=0$};
    \node at (-2.1,2.4) {$z=1$};
    \node at (3.6,0.3) {$ T = 0 $};
    \node at (3.6,2.3) {$  T = 0 $};
    
    \node at (0.7,-0.6) {${\color{colorbar15}\mathcal{F}_B }  $};
    \node at (0.7,3) {${\color{colorbar15}\mathcal{F}_T }  $}; 
    \node at (0.7,1.3) {$ H = 1$};
    
    \end{tikzpicture}
    \caption{ A non-dimensional schematic diagram for uniform internally heated convection. The upper and lower boundary is at the same temperature, the domain is periodic in the $x$ and $y$ directions, and $\mathcal{F}_B$ and $\mathcal{F}_T$ are the mean heat fluxes out the boundaries in the respective directions.  }
    \label{fig:config_ih1}
\end{figure}

IHC also has significant physical differences from RBC. In the turbulent regime, internal heating creates thermal boundary layers of different widths, with a flow characterised by plumes descending from an unstably stratified boundary layer at the top to a stably stratified one at the bottom \cite{Goluskin2016book}. 
Moreover, when the boundaries are isothermal, as shown in \cref{fig:config_ih1}, heat leaves through both boundaries, such that convection (and by extension $\wT$) leads to an asymmetry in the heat flux out of the domain \cite{goluskin2012convection}.
Indeed, in uniform IHC, we define
\begin{subequations}
\label{eq:Flux_def}
    \begin{align}
    \mathcal{F}_T &:= -\mean{\partial_zT}_h\vert_{z=1} = \frac12 + \wT, \\
    \mathcal{F}_B &:= \phantom{-}\mean{\partial_z T}_h\vert_{z=0} = \frac12 - \wT,
    \end{align}
\end{subequations}
as the non-dimensional mean heat fluxes through the top and bottom boundaries \cite{Goluskin2016book}. 
When the fluid is stationary, a state that is globally stable for $R\leq 26\,926.6$, then $\mean{wT}=0$ and all heat input is transported to the boundaries by conduction symmetrically, giving $\mathcal{F}_T = \mathcal{F}_B = \frac12$. Convection breaks this symmetry, causing more heat to escape through the top boundary. Given that the temperature remains non-negative in the domain, one can prove that $0 \leq \wT \leq \frac12$ uniformly in \Ra\ and \Pr~\cite{goluskin2012convection}. The zero lower bound of $\wT$ saturates for a no-flow state, saturating the upper bound of $\frac12$ would require a flow that transports heat upwards so efficiently that all heat escapes the domain through the top boundary.

Aided by numerical optimisation, the first $R$-dependent bounds on $\mathcal{F}_B$ were only recently proven\cite{Arslan2021,kumar2021ihc}. At arbitrary $Pr$, for all $R<R_n$ where $R_n$ is the Rayleigh number below which the use of the minimum principle on the temperature gives a suboptimal bound, $\mathcal{F}_B \gtrsim \frac12 - R^{1/5} $, while for $R>R_n$, instead $\mathcal{F}_B \gtrsim R^{1/5}\exp{(-R^{3/5})}$. At $Pr=\infty$ the best bound, prior to this work, was $\mathcal{F}_B \gtrsim R^{-2}$, notably the use different boundary layer widths at the top and bottom ensured no logarithmic corrections are present in the bound \cite{Arslan2023}. All this being, there remains room for improvement in bounding $\mathcal{F}_B$, as the bounds are conservative relative to known phenomenological theories and data from experiments and numerical simulations. Whether any flow saturates the known bounds on $\mathcal{F}_B$ remains an open question. The improvement in this work to the no-slip case shows that the choice of background fields used in Arslan et al.(2023) for what they called IH1 is suboptimal. Stated precisely the main contribution of this paper is the following two results,
\begin{subequations}
    \label{eq::results}
    \begin{align}
        \label{thm:ns}
          \textrm{no-slip:}&\qquad \mathcal{F}_B \geq  c_1 R^{-2/3} - c_2 R^{-1/2} \ln(1-c_3 R^{-1/3})  \qquad \forall R>R_0\, , \\
          \label{thm:sf}
            \textrm{stress-free:}&\qquad \mathcal{F}_B \geq  c_4 R^{-40/29} - c_5 R^{-35/29} \ln(1-c_6 R^{-10/29})  \qquad \forall R>R_0 \, .
    \end{align}
\end{subequations}
With positive constants $c_1$ to $c_6$ that are $O(1)$ and $R_0>1$.
One notable feature of the results in \eqref{thm:ns} and \eqref{thm:sf} is that in the limit of $R\rightarrow \infty $, the bounds are $c_1R^{-2/3}$ and $c_4R^{-40/29}$ respectively, however, for small $R$, the logarithmic term remains relevant.

For notation, we use $\lVert f \rVert_2$ to represent a standard $L^2$ norm of a function on $z\in(0,1)$, overbars to denote infinite-time averages, angled brackets to indicate volume averages, and angled brackets with a subscript $h$ for averages over only the horizontal directions:
\begin{subequations}
\begin{align}
    \horav{f} &= \frac{1}{L_x L_y} \int^{L_x}_0 \int^{L_y}_0  f(x,y,z,t) \, \textrm{d}y\, \textrm{d}x,
    \\
    \volav{f} &= 
    \int^{1}_{0} \horav{f} \,\textrm{d}z,\\[1ex]
    \mean{f} &= \limsup_{\tau\rightarrow \infty} \frac{1}{\tau} 
    \int^{\tau}_{0} \volav{f} \textrm{d}t.
\end{align}
\end{subequations}
The paper is organised as follows: \cref{sec:setup} presents the setup being considered, in \cref{sec:prob} by use of the auxiliary functional method, we construct the problem, \cref{sec:res} and \cref{sec:fs} prove the bounds for no-slip and stress-free boundary conditions of \cref{thm:ns} \& \cref{thm:sf}, then numerical results are in \cref{sec:num_res} for both boundary conditions and finally \cref{sec:conc} is a discussion of the bounds and numerical results with concluding remarks.

\section{Setup}
\label{sec:setup}
We consider an incompressible fluid with constant density $\rho$, specific heat capacity $c_p$ and thermal diffusivity $\kappa$ that is horizontally periodic between two plates a distance $d$ apart.
The fluid motion is governed by the Navier-Stokes equations under the Boussinesq approximation at infinite Prandtl number, heated at a constant rate, $H$, per unit volume. We take $d$ as the characteristic length scale, $d^2/ \kappa$ as the time scale and $ d^2 H/\kappa \rho c_p$ as the temperature scale \cite{roberts1967convection}. The fluid occupies the periodic domain $\Omega = [0,L_x] \times [0,L_x]   \times [0,1]$ and satisfies
\begin{subequations}\label{e:governing-equations}
    \begin{align}
    \nabla \cdot \vu &= 0\, , \label{continuit} \\
    \nabla p &= \Delta\vu + \Ra\, T \uvec{z}\, , \label{nondim_momentum} \\
    \partial_t T + \vu\cdot \nabla T  &= \Delta T + 1.
    \label{nondim_energy}
    \end{align}
\end{subequations}
where $\boldsymbol{u}=(u,v,w)$ is the velocity of the fluid in Cartesian coordinates, $p$ the pressure of the fluid and $T$ a scalar for the temperature of the fluid.
The only control parameter is a `flux' Rayleigh number defined as
\begin{equation}
    \Ra := \frac{g \alpha H d^{5}}{\rho c_p \nu \kappa^{2}}.
\end{equation}
Here $g$ is the acceleration of gravity, $\alpha$ is the thermal expansion coefficient of the fluid, and $\nu$ is the kinematic viscosity. At the boundaries, the velocity satisfies either no-slip or stress-free conditions, and the temperature is isothermal, taken as zero without loss of generality. Hence we enforce 
\begin{subequations}
\label{eq:bc}
\begin{align}
   \textrm{no-slip:}&\quad \vu|_{z\in\{0,1\}}=\boldsymbol{0}, 
    \label{bc_T_IH1} \\
    \textrm{stress-free:}&\quad   w|_{z\in\{0,1\}}= \partial_z u|_{z\in\{0,1\}} = \partial_z v|_{z\in\{0,1\}} =\boldsymbol{0}, 
    \label{bc_Tfs} \\
     &\quad T|_{z=0} =
    T|_{z=1} = 0.
    \label{bc_T}
\end{align}
\end{subequations}
To simplify the notation, we introduce a set that encodes the boundary conditions and the pointwise non-negativity constraint on the temperature,
\begin{subequations}
	\begin{gather}
	\label{e:T-space-ih1}
	\Tspace := \{(\boldsymbol{u},T) \, | \; \text{horizontal periodicity, $\nabla \cdot \boldsymbol{u}=0$ \&  \eqref{eq:bc} }  \},\\
	\Tspace_+ := \{ T \in \Tspace \,| \,T(\vx) \geq 0 \text{ a.e. } \vx \in \Omega \}.
	\label{e:T-positive-cone-ih1}
	\end{gather}
\end{subequations}

\section{The auxiliary function method}
\label{sec:prob}

Given that $\mathcal{F}_B$ is defined by \eqref{eq:Flux_def}, to bound $\mathcal{F}_B$, we find a bound on $\wT$, and will in the problem construction work with $\wT$. Here, we outline the main steps to make the paper self-contained, but further details are available in previous works \cite{Arslan2021,Arslan2023}.

To prove an upper bound on $\smash{\mean{wT}}$, we employ the auxiliary function method~\cite{Chernyshenko2014a,Fantuzzi2022}. The method relies on the observation that the time derivative of any bounded functional $\mathcal{V}\{T(t)\}$ along solutions of the Boussinesq equations~\eqref{e:governing-equations} averages to zero over infinite time, so that
\begin{equation}\label{e:af-method}
\mean{wT} = \timeav{\volav{wT} + \tfrac{\rm d}{ {\rm d} t}\mathcal{V}\{T(t)\}  }.
\end{equation}
Two key simplifications follow. The first is that we can estimate \eqref{e:af-method} by the pointwise-in-time maximum along the solutions of the governing equations, and then this value is estimated by the maximum it can take over \textit{all} velocity and temperature fields in $\mathcal{H}_+$.

We restrict our attention to quadratic functionals taking the form
\begin{equation}\label{e:V-IH1}
\mathcal{V}\{T\}: = \volav{ \frac{\bp}{2} |T|^2 - [\bfield(z)+z-1] T },
\end{equation}
that are parametrised by a positive constant $\bp > 0$, referred to as the balance parameter and a piecewise-differentiable function $\bfield:[0,1] \to \bR$ with a square-integrable derivative that we call the background temperature field.
Here $\bfield(z)$ satisfies
\begin{equation}\label{bc:psi}
\bfield(0)=1, \qquad \bfield(1)=0.
\end{equation}
Introducing a constant, $U$, and rearranging, \eqref{e:af-method} can be written as,
\begin{equation}
    \wT \leq U - U + \volav{ wT} + \tfrac{\textrm{d}}{\textrm{d}t}\mathcal{V}\{T\}   \leq U,
\end{equation}
where the final inequality holds given that, $ U - \volav{wT} - \tfrac{\textrm{d}}{\textrm{d}t}\mathcal{V}\{T\}  \geq0$. 
However, the minimum principle on $T$ is necessary to obtain a $R$-dependent bound on $\wT$ that approaches $\tfrac12$ from below as $R$ increases. The condition is enforced with a Lagrange multiplier\cite{Arslan2021,Arslan2023},  $\lambda(z)$, so that the problem statement becomes
\begin{align}
    \wT \leq \inf_{U,\beta,\tau(z),\lambda(z)} \{U ~|~ \mathcal{S}\{\boldsymbol{u},T\} \geq 0 \quad \forall(\boldsymbol{u},T)\in \Tspace_+ \} ,
    \label{eq:opt_prob_g}
\end{align}
provided $\lambda(z)$ is a non-decreasing function, where
\begin{equation}
    \label{e:S_new}
    \mathcal{S}\{\boldsymbol{u},T\} := \volav{\beta|\nabla T|^2 + \tau'(z) wT + (\beta z - \tau'(z) + \lambda(z))\partial_z T + \tau(z)+ U} - \tfrac12.
\end{equation}
From \eqref{eq:opt_prob_g}, an explicit expression on $U$ is obtained by exploiting horizontal periodicity and taking the Fourier decomposition,
\begin{equation}\label{e:Fourier}
    \begin{bmatrix}
    T(x,y,z)\\\boldsymbol{u}(x,y,z)
    \end{bmatrix}
    = \sum_{\boldsymbol{k}} 
    \begin{bmatrix}
    \hat{T}_{\boldsymbol{k}}(z)\\ \hat{\boldsymbol{u}}_{\boldsymbol{k}}(z)
    \end{bmatrix}
    \textrm{e}^{i(k_x x + k_y y)}\, ,
\end{equation}
where the sum is over wavevectors $\boldsymbol{k}=(k_x,k_y)$ for the horizontal periods $L_x$ and $L_y$. The magnitude of each wavevector is 
$k = \sqrt{k_{\smash{x}}^2 + k_{\smash{y}}^2}$.
Inserting the Fourier expansions~{(\ref{e:Fourier})} into~\eqref{e:S_new} and applying Youngs' inequality and using the incompressibility condition to write the horizontal Fourier amplitudes $\hat{u}_{\boldsymbol{k}}$ and $\hat{v}_{\boldsymbol{k}}$ in terms of $\hat{w}_{\boldsymbol{k}}$, gives an estimate from below on $\mathcal{S}\{\boldsymbol{u},T\}$. Then, using that $\hat{w}_0=0$ gives
\begin{equation}\label{e:S-estimate_n}
    \mathcal{S}\{\boldsymbol{u},T\}  \geq \mathcal{S}_{0}\{\hat{T}_0\} + \sum_{\boldsymbol{k}} \mathcal{S}_{\boldsymbol{k}} \{\hat{w}_{\boldsymbol{k}},\hat{T}_{\boldsymbol{k}}\},
\end{equation}
where
\begin{equation}
    \label{S0}
    \mathcal{S}_{0}\{\hat{T}_0\} := \int^{1}_0 \beta \vert \hat{T}_{0}' \vert^{2}+ (\beta z-\tau'(z)+\lambda(z))\hat{T}_{0}' + \tau(z) \,  \textrm{d}z + U - \frac12  ,
\end{equation}
and 
\begin{equation}\label{Sk}
    \mathcal{S}_{\boldsymbol{k}}\{\hat{w}_{\boldsymbol{k}},\hat{T}_{\boldsymbol{k}}\} := 
   \int^{1}_0 \beta \vert \hat{T}_{\boldsymbol{k}}' \vert^{2} + \beta k^{2}\vert \hat{T}_{\boldsymbol{k}} \vert^{2} 
    + \tau'(z)\textrm{Re}\{ \hat{w}_{\boldsymbol{k}}\hat{T}_{\boldsymbol{k}}^* \}\, \textrm{d}z \, , 
\end{equation}
with boundary conditions,
\begin{subequations}
\label{bc:fourier}
\begin{align}
    \label{bc:w_k}
    \textrm{no-slip:}&\quad \hat{w}_{\boldsymbol{k}}(0) =  \hat{w}_{\boldsymbol{k}}(1) =    \hat{w}'_{\boldsymbol{k}}(0) =  \hat{w}'_{\boldsymbol{k}}(1) = 0 ,\\
     \label{bc:w_k_fs}
    \textrm{stress-free:}&\quad \hat{w}_{\boldsymbol{k}}(0) =  \hat{w}_{\boldsymbol{k}}(1) =    \hat{w}''_{\boldsymbol{k}}(0) =  \hat{w}''_{\boldsymbol{k}}(1) = 0 ,\\
    \label{bc:T_k}& \quad
    \hat{T}_{\boldsymbol{k}}(0) =  \hat{T}_{\boldsymbol{k}}(1) = 0.
\end{align}
\end{subequations}
If $\mathcal{S}_0$ and $\mathcal{S}_{\boldsymbol{k}}$ independently are non-negative then $\mathcal{S}\geq 0$ is satisfied.   
The condition $\mathcal{S}_{\boldsymbol{k}} \geq 0$, is usually referred to as the \textit{spectral constraint} and must hold for all $k$, while guaranteeing  $\mathcal{S}_0\geq 0$ gives an explicit expression for $U$. Solving the Euler-Lagrange equations for $\hat{T}_0$ in \eqref{S0}, subject to the boundary conditions on $\tau(z)$ and $\hat{T}_0$ in \eqref{bc:psi} and \eqref{bc:T_k}, along with the condition $\int^{1}_0\lambda\textrm{d}z = -1$ gives,
\begin{equation}
    \label{eq:U}
    U := \frac12 + \frac{1}{4\beta} \Big\lVert \beta(z-\tfrac12) - \tau'(z) + \lambda(z) \Big\rVert_2^2 - \int^{1}_0 \tau(z)~  \textrm{d}z .
\end{equation}

The final ingredient is the diagnostic equation between the $w$ and $T$. Taking the vertical component of the double curl of the momentum equation \eqref{nondim_momentum} gives
\begin{equation}
\label{e:w_and_T_eq}
\Delta^2 w = -R \, \Delta_h T,
\end{equation}
where $\Delta_h := \partial^2_x + \partial^2_y $, is the horizontal Laplacian. Substituting for $w$ and $T$ given \eqref{e:Fourier} gives
\begin{equation}
\label{eq:w_T_ode}
    \hat{w}''''_{\boldsymbol{k}} - 2 k^2 \hat{w}''_{\boldsymbol{k}} + k^4 \hat{w}_{\boldsymbol{k}} = R k^2 \hat{T}_{\boldsymbol{k}} \, .
\end{equation}
Finally, the optimisation problem can be stated in a self-contained way as
\begin{equation}\label{e:optimization-Fourier}
    \begin{aligned}
        \inf_{\tau(z),\lambda(z),\beta} \quad &\frac12 +  \frac{1}{4\beta} \Big\lVert \beta(z-\tfrac12) - \tau'(z) + \lambda(z) \Big\rVert_2^2 - \int^{1}_0 \tau(z)~  \textrm{d}z  , \\
        \text{subject to} \quad
        &\tau(0)=1, \tau(1)=0, \\ 
        &  \volav{\lambda(z)}=-1~~ \& ~~\lambda(z) ~\textrm{non-decreasing},
        \\
        &\mathcal{S}_{\boldsymbol{k}}\{ \hat{w}_{\boldsymbol{k}},\hat{T}_{\boldsymbol{k}}\} \geq 0 \qquad\forall \hat{w}_{\boldsymbol{k}},\hat{T}_{\boldsymbol{k}}\,|\, \eqref{bc:T_k}, \eqref{bc:w_k}\, \textrm{or}\, \eqref{bc:w_k_fs} \quad \forall \boldsymbol{k}\neq 0, \\
        &\hat{w}''''_{\boldsymbol{k}} - 2 k^2 \hat{w}''_{\boldsymbol{k}} + k^4 \hat{w}_{\boldsymbol{k}} = R k^2 \hat{T}_{\boldsymbol{k}}.
    \end{aligned}
\end{equation}

\section{Bounds for no-slip boundaries}
\label{sec:res}

In this chapter, we prove an upper bound on $\wT$ with the auxiliary function method under no-slip boundary conditions. For the problem constructed in \cref{sec:prob}, we first introduce choices for the background field $\tau(z)$, Lagrange multiplier $\lambda(z)$ and balance parameter $\beta$ in \cref{sec:prelim}. Then, we utilise an estimate on $\hat{w}''_{\boldsymbol{k}}$, first proven in Doering \& Constantin (2002) \cite{Doering2001}, that uses the diagnostic equation \eqref{eq:w_T_ode}, to enforce the spectral constraint. We do not attempt to optimise the constants and in \cref{sec:R_bound} prove the desired result.

\subsection{Preliminaries}\label{ss:ansatz-IH1}
\label{sec:prelim}

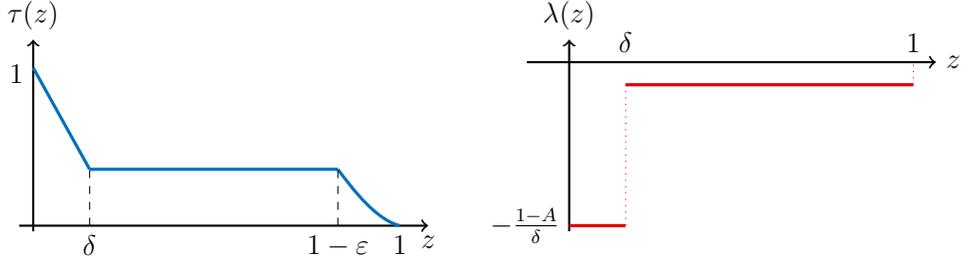
\begin{figure}
	\centering
    \begin{tikzpicture}[every node/.style={scale=0.9}, scale=0.75]
	\draw[->,black,thick] (-12.25,0.5) -- (-5,0.5) node [anchor=north] {$z$};
	\draw[->,black,thick] (-12,0.4) -- (-12,3.8) node [anchor=south] {$\tau(z)$};
	\draw[matlabblue,very thick] (-12,3.3) -- (-11,1.5) ;
    \draw[matlabblue,very thick] (-11,1.5) -- (-6.6,1.5);
    \draw[matlabblue, very thick] plot [smooth, tension = 1] coordinates {(-6.6,1.5) (-6,0.8) (-5.5,0.5)};
	\node[anchor=north] at (-5.5,0.5) {$1$};
	\node[anchor=east] at (-12,3.2) {$1$};
    \draw[dashed] (-11,0.5) node[anchor=north] {$\delta$} -- (-11,1.5);
	\draw[dashed] (-6.6,0.5) node[anchor=north] {$1-\varepsilon$} -- (-6.6,1.5);
	\draw[->,black,thick] (-3.25,3.4) -- (4,3.4) node [anchor=west] {$z$};
	\draw[->,black,thick] (-2.5,0.4) -- (-2.5,3.8) node [anchor=south] {$\lm(z)$};
	\draw[colorbar12,very thick] (-2.5,0.5) -- (-1.5,0.5);
	\draw[colorbar12,very thick] (-1.5,3) -- (3.6,3) ;
	\draw[colorbar12,dotted] (-1.5,3) -- (-1.5,0.5) ;
 \draw[colorbar12,dotted] (3.6,3) -- (3.6,3.4) ;
	\node[anchor=south] at (3.6,3.4) {$1$};
	\node[anchor=south] at (-1.5,3.4) {$\delta$};
	\node[anchor=east] at (-2.5,0.5) {$-\frac{1-A}{\delta} $};
    \end{tikzpicture}
	\caption{Sketches of the functions $\bfield(z)$ in~\eqref{eq:tau_c} and $\lm(z)$ in~\eqref{e:q_profile} used to prove \eqref{thm:ns}, where $\delta$ is the boundary layer width at the bottom, $\varepsilon$ the boundary layer width at the top of the domain and $A$ is given by \cref{e:A-choice}.}
	\label{fig:psi_inf_pr}
\end{figure}

To prove the upper bound on $\wT$ requires appropriate choices of $\bp>0$, $\bfield(z)$, and $\lm(z)$ that satisfy the conditions of \eqref{e:optimization-Fourier} and make the quantity $U(\bp,\bfield,\lm)$ as small as possible. To simplify this task, given the conditions that $\tau(0)=1$, $\tau(1)=0$ and $\volav{\lambda}=-1$, we restrict $\bfield(z)$ to take the form
\begin{equation}
    \label{eq:tau_c}
    \tau(z) = 
    \begin{dcases}
        1-\frac{(1-A) z}{ \delta },& 0\leq z\leq \delta ,\\
         A, & \delta \leq z \leq 1- \varepsilon, \\
        \frac{A}{\varepsilon} (1-\varepsilon) \left(\frac{1-z}{z}\right), & 1-\varepsilon\leq z \leq 1, 
    \end{dcases}
\end{equation}
and $\lm(z)$ to be given by
\begin{equation}
\label{e:q_profile}
\lm(z):= 
\begin{dcases}
-\frac{1-A}{\delta} ,&  0\leq z \leq \delta,\\
-\frac{A}{1-\delta}, & \delta \leq z \leq 1.
\end{dcases}
\end{equation}
These piecewise-defined functions, sketched in \cref{fig:psi_inf_pr}, are fully specified by the bottom boundary layer width $\delta \in (0,\frac13)$, the top boundary layer width $\varepsilon \in (0,\frac13)$, and the parameter $A > 0$ that determines the amplitude of $\bfield(z)$ in the bulk of the layer. The upper limit of $\frac13$ on the boundary layers' widths is set here for the convenience of the algebra. Smaller maximal values of $\varepsilon$ and $\delta$ will give a bound with improved prefactors without changing the $R$ scaling.

We also fix
\begin{equation}\label{e:A-choice}
A = \frac{1}{\sqrt{2}} \delta \varepsilon^{1/2}.
\end{equation}
This choice arises when insisting that the upper bound on $U$ be strictly less than $\frac12$ for values of $\delta$ and $\varepsilon$. From a similar argument, the sign-positive integral in the expression of $U$ in \eqref{eq:U} can be estimated as
\begin{equation}
\label{eq:tri_eq}
\frac{1}{4\bp} \left\lVert  \bfield'(z) - \lm(z) - \bp \left(z - \tfrac{1}{2}\right)  \right\rVert_2^2 \leq \frac{\bp}{2} \lVert z - \tfrac{1}{2}\rVert_2^2 + \frac{1}{2\bp} \lVert \bfield'(z)- \lm(z) \rVert_2^2\, .
\end{equation}
Hence, we also fix 
\begin{equation}\label{e:b-choice}
\bp :=  \lVert z - \tfrac12 \rVert_2^{-1}\lVert \bfield'(z)- \lm(z) \rVert_2  = 2\sqrt{3} \lVert \bfield'(z) - \lm(z) \rVert_2 \, .
\end{equation}

Finally, we require the following result from Doering and Constaint (2001) \cite{Doering2001} that provides a pointwise estimate on the magnitude of $\hat{w}''_{\boldsymbol{k}}$ in terms of $\hat{T}_{\boldsymbol{k}} $.
\begin{lemma}[Velocity estimate \cite{Doering2001}]
    \label{lem:DC_inf}
    Let $T,w : \Omega \rightarrow \mathbb{R}$, be horizontally periodic functions such that $\Delta^2 w = - R \Delta_h T$ subject to the velocity conditions in \eqref{bc_T_IH1}. Then
    \begin{equation}
        \label{eq:w_est_m}
        \lVert \hat{w}''_{\boldsymbol{k}} \rVert_\infty \leq c_0 R k \lVert \hat{T}_{\boldsymbol{k}} \rVert_2 \,,
    \end{equation}
    where $c_0 = \sqrt{2/(7-\sqrt{41})}$ .
\end{lemma}

\subsection{ Estimates on the upper bound }
\label{sec:R_bound}

Given the choices stated in \cref{sec:prelim} we first obtain an upper bound on $U$ in terms of the lower boundary layer $\delta$ in $\tau(z)$ given by \eqref{eq:tau_c}.

To start off, use of \eqref{e:b-choice} and \eqref{eq:tri_eq} in the expression of $U$ in \eqref{eq:U} gives,  
\begin{equation}
    \label{eq:U_ineq}
    U \leq \frac12 + \frac{1}{\sqrt{12}} \lVert \tau'(z) - \lambda(z)\rVert_2 - \int^{1}_0 \tau(z) ~\textrm{d}z .
\end{equation}
Next, we evaluate the two integrals in \eqref{eq:U_ineq}. The positive-definite integral in \eqref{eq:U_ineq} is estimated from above and below, as both are necessary for the proof. Given  $\tau(z)$ and $\lambda(z)$ in \eqref{eq:tau_c} and \eqref{e:q_profile} we have, 
\begin{align}
   \lVert \tau'(z)-\lambda(z) \rVert_2^2 &= \int^{1}_\delta |\tau'(z)-\lambda(z)|^2 \textrm{d}z\nonumber\\ & = \frac{A^2}{(1-\delta)^2}(1-\varepsilon-\delta)  + \frac{A^2}{\varepsilon^2} \int^{1}_{1-\varepsilon} \left(-\frac{(1-\varepsilon)}{z^{2}} + \frac{\varepsilon}{1-\delta}\right)^2 \textrm{d}z  \nonumber \\ 
    & =  \frac{A^2}{3\varepsilon}\frac{(3-3\varepsilon+\varepsilon^2)}{1-\varepsilon} - \frac{A^2}{1-\delta}.
    \label{eq:2norm_exac}
\end{align}
To obtain a lower bound on \eqref{eq:2norm_exac}, the non-negativity of $\varepsilon$ gives $1/(1-\varepsilon) \geq 1$ and given that $\varepsilon,\delta\leq \frac13$, we take $3-3\varepsilon+\varepsilon^2\geq \frac{19}{9} $ and $-(1-\delta)^{-1}\geq -( \frac{17}{27} ) \varepsilon^{-1}$ to obtain
\begin{equation}
\label{eq:lower_U}
   \lVert \tau'(z)-\lambda(z) \rVert_2^2 \geq \frac{2A^2}{27\varepsilon}. 
\end{equation}
For an upper bound on \eqref{eq:2norm_exac}, given that $\varepsilon\leq \frac13$ we take $3-3\varepsilon+\varepsilon^2 \leq 4$ and $1/(1-\varepsilon)\leq 2$, such that 
\begin{equation}
\label{eq:upper_U}
      \lVert \tau'(z)-\lambda(z) \rVert_2^2 \leq \frac{8A^2}{3\varepsilon}. 
\end{equation}
Then, when we evaluate the integral of $\tau(z)$ in \eqref{eq:U} to obtain 
\begin{equation}
    \label{eq:tau_int}
   \int^{1}_0 \tau(z)~ \textrm{d}z = \frac12 \delta(1-A) - \frac{A}{\varepsilon}(1-\varepsilon)\ln(1-\varepsilon).
\end{equation}
Substituting \eqref{eq:tau_int} and \eqref{eq:upper_U} back into \eqref{eq:U_ineq}, taking $A$ as given by \eqref{e:A-choice} and $\varepsilon,\delta\leq \frac13$ such that $\delta^2\varepsilon^{1/2} \leq  (\frac{\sqrt{2}}{6} )\delta$ and $-(1-\varepsilon)\leq -\frac23$, gives
\begin{equation}
    U \leq \frac12 -\frac{1}{12} \delta  + \frac{2}{3\sqrt{2}} \delta \varepsilon^{-1/2} \ln{(1-\varepsilon)} \, .
    \label{eq:U_fa}
\end{equation}

\subsection{Satisfying the spectral constraint}

Next, we determine the non-negativity of the spectral constraint, $\mathcal{S}_{\boldsymbol{k}}\geq 0$. Taking the absolute magnitude of the sign-indefinite integral in \eqref{Sk} and substituting for $\tau(z)$ from \eqref{eq:tau_c} gives
\begin{equation}
    \label{eq:sign_indef_sk}
    \int^{1}_0\tau' |\hat{w}_{\boldsymbol{k}} \hat{T}_{\boldsymbol{k}} | \textrm{d}z = -\frac{1-A}{\delta}\int^{\delta}_0 |\hat{w}_{\boldsymbol{k}} \hat{T}_{\boldsymbol{k}} |\textrm{d}z - \frac{A}{\varepsilon}(1-\varepsilon) \int^{1}_{1-\varepsilon} z^{-2} |\hat{w}_{\boldsymbol{k}} \hat{T}_{\boldsymbol{k}} |\textrm{d}z .
\end{equation}
First, we estimate the integral at the lower boundary in \eqref{eq:sign_indef_sk}. Given the boundary conditions in \eqref{bc:fourier}, use of the fundamental theorem of calculus and H\"olders inequality gives
\begin{equation}
    \label{eq:w_est}
    |\hat{w}_{\boldsymbol{k}}| = \int^{z}_{0} \int^{\sigma}_{0} |\partial^2_\eta \hat{w}_{\boldsymbol{k}}(\eta)| \textrm{d}\eta \textrm{d}\sigma \leq \frac12 z^2 \lVert \hat{w}''_{\boldsymbol{k}} \rVert_{\infty},
\end{equation}
and for $\hat{T}_{\boldsymbol{k}}$, the fundamental theorem of calculus and the Cauchy-Schwarz inequality give 
\begin{equation}
    \label{eq:T_est}
    |\hat{T}_{\boldsymbol{k}}| = \int^{z}_{0} |\partial_\eta \hat{T}_{\boldsymbol{k}}(\eta)| \textrm{d}\eta  \leq \sqrt{z} \lVert \hat{T}'_{\boldsymbol{k}} \rVert_{2}.
\end{equation}
Then, by use of \eqref{eq:w_est} and \eqref{eq:T_est}, the integral at the lower boundary, given that $0<A<1$, becomes
\begin{align}
   \frac{1-A}{\delta} \int^{\delta}_0  |\hat{w}_{\boldsymbol{k}} \hat{T}_{\boldsymbol{k}} | \textrm{d}z \leq \frac{1}{\delta} \int^{\delta}_0  |\hat{w}_{\boldsymbol{k}} |  |\hat{T}_{\boldsymbol{k}} | \textrm{d}z &\leq \frac{1}{2\delta}\int^{\delta}_0  z^{5/2}  \textrm{d}z \lVert \hat{w}''_{\boldsymbol{k}} \rVert_{\infty } \lVert \hat{T}'_{\boldsymbol{k}} \rVert_{2} \nonumber\\
   &= \frac17 \delta^{5/2}  \lVert \hat{w}''_{\boldsymbol{k}} \rVert_{\infty } \lVert \hat{T}'_{\boldsymbol{k}} \rVert_{2}.
   \label{eq:low_est_f}
\end{align}
Now we use \cref{lem:DC_inf}, substituting \eqref{eq:w_est_m} for $\hat{w}''_{\boldsymbol{k}}$ in \eqref{eq:low_est_f} and using Youngs' inequality gives
\begin{equation}
    \frac17 \delta^{5/2}  \lVert \hat{w}''_{\boldsymbol{k}} \rVert_{\infty} \lVert \hat{T}'_{\boldsymbol{k}} \rVert_{2} \leq \frac17 c_0 k \delta^{5/2} R   \lVert \hat{T}_{\boldsymbol{k}} \rVert_{2}  \lVert \hat{T}'_{\boldsymbol{k}} \rVert_{2} \leq \frac{\beta}{2} k^2 \lVert \hat{T}_{\boldsymbol{k}} \rVert_{2}^2 + \frac{c_0^2 }{98} \frac{ \delta^5 R^2}{\beta}   \lVert \hat{T}'_{\boldsymbol{k}} \rVert_{2}^2 \, . 
    \label{eq:lower_est}
\end{equation}
Turning to the integral at the upper boundary, we modify \eqref{eq:w_est} and \eqref{eq:T_est} into
\begin{equation}
    \label{eq:w_T_est_u}
     |\hat{w}_{\boldsymbol{k}}|  \leq \frac12 (1-z)^2 \lVert \hat{w}''_{\boldsymbol{k}} \rVert_{\infty}, \quad  \textrm{and} \quad |\hat{T}_{\boldsymbol{k}}|   \leq \sqrt{1-z}\, \lVert \hat{T}'_{\boldsymbol{k}} \rVert_{2}.
\end{equation}
Given that the integral is evaluated in the interval $(1-\varepsilon,1)$, where $\varepsilon\leq \frac13$, we have that $z^{-2} \leq (1-\varepsilon)^{-2} \leq \frac52 $ and $1-\varepsilon\leq 1$. Following the same steps as before, using \eqref{eq:w_est_m} and Youngs' inequality gives
\begin{align}
    \frac{A(1-\varepsilon)}{\varepsilon}\int^{1}_{1-\varepsilon} \frac{ |\hat{w}_{\boldsymbol{k}} \hat{T}_{\boldsymbol{k}}|}{z^2} \textrm{d}z &\leq  \frac{A}{\varepsilon}\int^{1}_{1-\varepsilon} \frac{ |\hat{w}_{\boldsymbol{k}} \hat{T}_{\boldsymbol{k}}|}{z^2}  \textrm{d}z \leq \frac{A}{2\varepsilon} \int^{1}_{1-\varepsilon} \frac{(1-z)^{5/2}}{z^2} \textrm{d}z \lVert \hat{w}''_{\boldsymbol{k}} \rVert_{\infty} \lVert \hat{T}'_{\boldsymbol{k}}\rVert_2 \nonumber \\
    & \leq \frac{A}{2\varepsilon} \int^{1}_{1-\varepsilon} \frac{(1-z)^{5/2}}{(1-\varepsilon)^2} \textrm{d}z \lVert \hat{w}''_{\boldsymbol{k}} \rVert_{\infty } \lVert \hat{T}'_{\boldsymbol{k}}\rVert_2  \nonumber\\
    &\leq \frac{c_0 k A\varepsilon^{5/2}  R }{7(1-\varepsilon)^2}  \lVert \hat{T}_{\boldsymbol{k}}\rVert_2 \lVert \hat{T}'_{\boldsymbol{k}}\rVert_2  \leq \frac{5}{14} c_0 k A\varepsilon^{5/2} R  \lVert \hat{T}_{\boldsymbol{k}}\rVert_2 \lVert \hat{T}'_{\boldsymbol{k}}\rVert_2\nonumber\\
    &\leq \frac{\beta}{2} k^2 \lVert \hat{T}_{\boldsymbol{k}}\rVert_2^2+ \frac{25c_0^2 }{392} \frac{A^2 \varepsilon^5 R^2}{\beta} \lVert \hat{T}'_{\boldsymbol{k}}\rVert_2^2.
    \label{eq:upper_est}
\end{align}
Substituting \eqref{eq:lower_est} and \eqref{eq:upper_est} into \eqref{eq:sign_indef_sk} gives that
\begin{equation}
    \int^{1}_0 \tau' |\hat{w}_{\boldsymbol{k}}\hat{T}_{\boldsymbol{k}}| \textrm{d}z \geq - \beta k^2 \lVert \hat{T}_{\boldsymbol{k}} \rVert_2^2 - \left( \frac{c_0^2 }{98} \frac{ \delta^5 R^2}{\beta}  + \frac{25c_0^2 }{392} \frac{A^2 \varepsilon^5 R^2}{\beta} \right) \lVert \hat{T}'_{\boldsymbol{k}} \rVert_2^2.
    \label{eq:sign_indef1}
\end{equation}
After substituting \eqref{eq:sign_indef1} into \eqref{Sk} we get
\begin{equation}
    \label{eq:sk_f}
    \mathcal{S}_{\boldsymbol{k}}\{\hat{w}_{\boldsymbol{k}},\hat{T}_{\boldsymbol{k}}\} \geq \left(\beta - \frac{c_0^2 }{98} \frac{ \delta^5 R^2}{\beta}  - \frac{25c_0^2 }{392} \frac{A^2 \varepsilon^5 R^2}{\beta} \right) \lVert \hat{T}'_{\boldsymbol{k}} \rVert_2^2 \geq 0.
\end{equation}
Finally, given the estimates in \eqref{eq:lower_U} and \eqref{eq:upper_U} we have estimates for $\beta$ from above and below, where $\beta$ is given by \cref{e:b-choice} such that
\begin{equation}
    \label{eq:beta_est}
   \frac23 \delta \leq \beta \leq 4 \delta. 
\end{equation}
Use of the lower bounds on $\beta$ from \eqref{eq:beta_est} and substituting for $A$ into \eqref{eq:sk_f} the spectral constraint is satisfied provided 
\begin{equation}
    1 - \frac{9c_0^2}{ 392 }  \delta^3 R^2 - \frac{225c_0^2 }{3136} \varepsilon^6 R^2 \geq 0. 
\end{equation}
We then make the choice $\delta = \frac{25^{1/3}}{2} \varepsilon^2$, and take $c_0$ as given by \cref{lem:DC_inf} such that 
\begin{equation}
    \label{eq:delta}
    \delta \leq \left(\frac{98}{9}(7-\sqrt{41})\right)^{1/3} R^{-2/3},
\end{equation}
is the condition necessary to ensure that the spectral constraint is satisfied.

\subsection{Bound on \texorpdfstring{$\mathcal{F}_B$}{test 1}}

Taking $\delta$ as large as possible in \eqref{eq:delta} gives the condition with the largest freedom that satisfies the spectral constraint. Finally, substituting back into the upper bound on $U$ in \cref{eq:U_fa}, remembering that $\wT\leq U$, gives
\begin{equation}
    \wT \leq \frac12 - n_1 R^{-2/3} + n_2 R^{-1/2}  \ln{(1- n_3 R^{-1/3})} \, ,
\end{equation}
where $n_1 = \frac{1}{12} \left(\frac{98}{9}(7-\sqrt{41})\right)^{1/3} = 0.1555$, $n_2 = \frac{2^{1/6}}{3} n_3^{3/2} = 0.7104 $ and $n_3 = 2^{1/6}(12 n_1)^{1/2} = 1.5334 $ . By \eqref{eq:Flux_def}, we obtain that
\begin{equation}
    \mathcal{F}_B \geq n_1 R^{-2/3} - n_2 R^{-1/2}  \ln{(1- n_3 R^{-1/3})},
\end{equation}
the desired final bound.

\section{Bound for stress-free boundaries}
\label{sec:fs}

For stress-free boundary conditions, the optimisation problem in \eqref{e:optimization-Fourier} is identical. 
In what follows, we will demonstrate an upper bound on $\wT$ to bound $\mathcal{F}_B$, with a new background field, $\tau(z)$ and Lagrange multiplier, $\lambda(z)$. Then, by a pseudo-vorticity, $\zeta$, first used by Whitehead and Doering (2012)\cite{Whitehead2012}, we demonstrate that the spectral constraints is satisfied and  again choose not to optimise constants.

\subsection{Preliminaries}

\begin{figure}
	\centering
    \begin{tikzpicture}[every node/.style={scale=0.9}, scale=0.75]
	\draw[->,black,thick] (-12.25,0.5) -- (-5,0.5) node [anchor=north] {$z$};
	\draw[->,black,thick] (-12,0.4) -- (-12,3.8) node [anchor=south] {$\tau(z)$};
	\draw[matlabblue,very thick] (-12,3.3) -- (-11,0.5) ;
    \draw[matlabblue,very thick] (-11,0.5) -- (-6.9,1.5);
   \draw[matlabblue, very thick] plot [smooth, tension = 1] coordinates {(-6.9,1.5) (-6.2,0.85) (-5.5,0.5)};
	\node[anchor=north] at (-5.5,0.5) {$1$};
	\node[anchor=east] at (-12,3.2) {$1$};
    \node[anchor=north] at (-11,0.5) {$\delta$};
	\draw[dashed] (-6.9,0.5) node[anchor=north] {$1-\varepsilon$} -- (-6.9,1.4);
	\draw[->,black,thick] (-3.25,3.4) -- (4,3.4) node [anchor=west] {$z$};
	\draw[->,black,thick] (-2.5,0.4) -- (-2.5,3.8) node [anchor=south] {$\lm(z)$};
	\draw[colorbar12,very thick] (-2.5,0.5) -- (-1.5,0.5);
	\draw[colorbar12,very thick] (-1.5,3.4) -- (3.6,3.4) ;
	\draw[colorbar12,dotted] (-1.5,3.4) -- (-1.5,0.5) ;
	\node[anchor=south] at (3.6,3.4) {$1$};
	\node[anchor=south] at (-1.5,3.4) {$\delta$};
	\node[anchor=east] at (-2.5,0.5) {$-\frac{1}{\delta} $};
    \end{tikzpicture}
	\caption{Sketches of the functions $\bfield(z)$ in~\eqref{eq:tau_fs} and $\lm(z)$ in~\eqref{e:q_fs} used to prove \eqref{thm:sf}, where $\delta$ is the boundary layer width at the bottom and $\varepsilon$ the boundary layer width at the top of the domain.}
	\label{fig:psi_inf_pr_fs}
\end{figure}
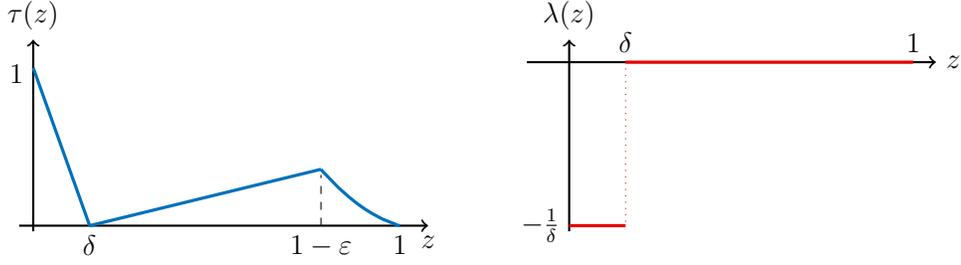

Similar to \cref{sec:res}, we initially state preliminary choices and estimates to obtain an upper bound on $\wT$, with some $\beta>0$, $\tau(z)$ and $\lambda(z)$ that satisfy the conditions in \eqref{e:optimization-Fourier} subject to stress-free boundary conditions. We choose $\tau(z)$ to be,
\begin{equation}
    \label{eq:tau_fs}
    \tau(z) = 
    \begin{dcases}
        1-\frac{z}{ \delta },& 0\leq z\leq \delta ,\\
         A(z-\delta), & \delta \leq z \leq 1- \varepsilon, \\
        \frac{A}{\varepsilon} (1-\varepsilon-\delta)(1-\varepsilon) \left(\frac{1-z}{z}\right), & 1-\varepsilon\leq z \leq 1, 
    \end{dcases}
\end{equation}
and $\lm(z)$ to be given by
\begin{equation}
\label{e:q_fs}
\lm(z):= 
\begin{dcases}
-\frac{1}{\delta} ,&  0\leq z \leq \delta,\\
0, & \delta \leq z \leq 1.
\end{dcases}
\end{equation}
These piecewise-defined functions, sketched in \cref{fig:psi_inf_pr_fs}, are fully specified by the bottom boundary layer width $\delta \in (0,\frac13)$, the top boundary layer width $\varepsilon \in (0,\frac13)$, and the parameter $A > 0$.
We fix $A$ to be 
\begin{equation}
    A = \frac{\delta \varepsilon^{1/2}}{3},
    \label{eq:A_2}
\end{equation}
but take $\beta$ as given by \eqref{e:b-choice}.

For stress-free boundary conditions, we introduce the pseudo-vorticity $\zeta$, which, due to horizontal periodicity and incompressibility in Fourier space, is given by,
\begin{equation}
    \label{eq:psu_vort}
    k \hat{\zeta}_{\boldsymbol{k}} = \hat{w}''_{\boldsymbol{k}} - k^2 \hat{w}_{\boldsymbol{k}} ,
\end{equation}
with boundary conditions
\begin{equation}
    \label{bc:p_v_ff}
    \hat{\zeta}_{\boldsymbol{k}}(0) =  \hat{\zeta}_{\boldsymbol{k}}(1) = 0.
\end{equation}
Use of \eqref{eq:psu_vort} along with incompressibility and \eqref{eq:w_T_ode} gives the following lemma.
\begin{lemma}[Pseudo-vorticity and integral estimates\cite{Whitehead2012}]
\label{lem:inf_fs}
Let $T,\boldsymbol{u}: \Omega \rightarrow \mathbb{R}$ be horizontally periodic functions such that $\Delta^2 w = -R \Delta_h T $ subject to the velocity boundary conditions. Let the pseudo-vorticity $\hat{\zeta}_{\boldsymbol{k}}$ be given by \eqref{eq:psu_vort}.Then,
\begin{equation}
    \label{eq:T_omega_lem}
    k^2 R^2 \lVert \hat{T}_{\boldsymbol{k}}\rVert_2^2 \geq k^4 \lVert \hat{\zeta}_{\boldsymbol{k}}\rVert_2^2 ,
\end{equation}
and
\begin{equation}
    \label{eq:wt_omega_lem}
    \int^{1}_0|\hat{w}_{\boldsymbol{k}}\hat{T}_{\boldsymbol{k}}|\textrm{d}z = \frac{1}{R} \lVert \hat{\zeta}_{\boldsymbol{k}}\rVert_2^2 .
\end{equation}
\end{lemma}
Finally, we require an additional pointwise estimate of the vertical velocity by the pseudo-vorticity. The estimate first introduced for RBC for the 2D scalar vorticity  also applies to the pseudo-vorticity in 3D.
\begin{lemma}[Pointwise velocity estimate\cite{Whitehead2011prl}]
\label{lem:w_vort_est}
    Let $\hat{w}_{\boldsymbol{k}}$ and $\hat{\zeta}_{\boldsymbol{k}}$ be functions zero at $z=0$ and $z=1$. Given the relation \eqref{eq:psu_vort} and incompressibility $\nabla \cdot \boldsymbol{u} = 0$, it follows that
    \begin{equation}
        \label{eq:w_vort_est}
        |\hat{w}_{\boldsymbol{k}}| \leq c_1\,  k^{\frac12} \min{(z,1-z)} \lVert \hat{\zeta}_{\boldsymbol{k}}\rVert_{2},
    \end{equation}
    where $c_1 = 3^{3/4} / 2^{3/2}$.
\end{lemma}

\subsection{Estimates on the upper bound}

We take the same choice of $\beta$ as in the previous section, and the expression of $U$ is estimated from above by \eqref{eq:U_ineq}. Then, given $\tau(z)$ and $\lambda(z)$ in  \eqref{eq:tau_fs} and \eqref{e:q_fs} we get
\begin{align}
    \lVert \tau'(z) - \lambda(z) \rVert_2^2 &= \int^{1}_\delta |\tau'|^2 \textrm{d}z  = A^2(1-\varepsilon-\delta) + \frac{A^2}{\varepsilon^2}(1-\varepsilon-\delta)^2 (1-\varepsilon)^2 \int^{1}_{1-\varepsilon}\frac{1}{z^{4}} \textrm{d}z \nonumber \\
     &= A^2(1-\varepsilon-\delta) + \frac{A^2}{3\varepsilon} \frac{(1-\varepsilon-\delta)^2(3-3\varepsilon+\varepsilon^2)}{1-\varepsilon}.
     \label{eq:l2_explicit}
\end{align}
For a lower bound on \eqref{eq:l2_explicit}, the fact that $\varepsilon\leq \frac13 $ and $\delta\leq \frac13$, implies that $1-\varepsilon-\delta\geq \frac13$, $1/(1-\varepsilon)\geq 1$ and $3-3\varepsilon+\varepsilon^2 \geq 2$ such that
\begin{equation}
\label{eq:B_sf_low}
    \lVert \tau'(z) - \lambda(z) \rVert_2^2 \geq \frac{2 A^2}{27 \varepsilon}.
\end{equation}
Whereas for an upper bound, $1-\varepsilon-\delta\leq 1$, $3-3\varepsilon+\varepsilon^2 \leq 4$ and $1/(1-\varepsilon)\leq 2$ gives
\begin{equation}
    \label{eq:2norm_e}
       \lVert \tau'(z) - \lambda(z) \rVert_2^2 \leq \frac{8 A^2}{3 \varepsilon} + A^2 \leq \frac{3A^2}{\varepsilon}.
\end{equation}
The integral of the background field is
\begin{equation}
    \label{eq:int_psi}
    \int^{1}_0 \tau(z) ~ \textrm{d}z = \frac12 \delta - \frac12 A (1-\varepsilon-\delta)(1-\varepsilon+\delta) - \frac{A}{\varepsilon}(1-\varepsilon)(1-\varepsilon- \delta)\ln{(1-\varepsilon)}.
\end{equation}
Then, substituting \eqref{eq:int_psi} and \eqref{eq:2norm_e} back into \eqref{eq:U_ineq} with $A$ as given by \eqref{eq:A_2}, we obtain after use of the estimate $ \delta\varepsilon^{1/2} \leq  \delta$, that
\begin{equation}
    U \leq \frac12 -\frac{1}{6} \delta  + \frac{2}{27} \delta \varepsilon^{-1/2} \ln{(1-\varepsilon)} \, .
    \label{eq:U_fas}
\end{equation}

\subsection{Enforcing the spectral constraint}

Now we find the condition for the spectral constraint, $\mathcal{S}_{\boldsymbol{k}}\geq0$, to hold. Since $\varepsilon,\delta\leq \frac13$ we have the estimate $(1-\varepsilon-\delta)(1-\varepsilon)\leq 1$. Substituting for $\tau'(z)$ from \eqref{eq:tau_fs}, $A$ from \eqref{eq:A_2}, use of \eqref{eq:T_omega_lem} and \eqref{eq:wt_omega_lem} from \cref{lem:inf_fs} in \eqref{Sk} and rearranging gives 
\begin{align}
    \mathcal{S}_{\boldsymbol{k}}\{ \hat{w}_{\boldsymbol{k}}, \hat{T}_{\boldsymbol{k}} \} &\geq \beta \lVert \hat{T}'_{\boldsymbol{k}} \rVert_2^2 + \frac{\beta k^4}{R^2} \lVert \hat{\zeta}_{\boldsymbol{k}} \rVert_2^2 -\frac{1}{\delta} \int^{\delta}_0 |\hat{w}_{\boldsymbol{k}} \hat{T}_{\boldsymbol{k}}|\textrm{d}z + A \int^{1-\varepsilon}_\delta |\hat{w}_{\boldsymbol{k}} \hat{T}_{\boldsymbol{k}}|\textrm{d}z \nonumber \\ 
    & \qquad\qquad\quad - \frac{A}{\varepsilon}(1-\varepsilon-\delta)(1-\varepsilon) \int^{1}_{1-\varepsilon} z^{-2} |\hat{w}_{\boldsymbol{k}} \hat{T}_{\boldsymbol{k}}|\textrm{d}z \\
     & \geq \beta \lVert \hat{T}'_{\boldsymbol{k}} \rVert_2^2 + \left( \frac{\beta k^4}{R^2} + \frac{\delta \varepsilon^{1/2}}{3 R} \right) \lVert \hat{\zeta}_{\boldsymbol{k}} \rVert_2^2 - \left(\frac{1}{\delta} + \frac{\delta \varepsilon^{1/2}}{3} \right) \int^{\delta}_0 |\hat{w}_{\boldsymbol{k}} \hat{T}_{\boldsymbol{k}}|\textrm{d}z  \nonumber \\ 
    & \qquad\qquad\quad - \frac{\delta}{3\, \varepsilon^{1/2}} \int^{1}_{1-\varepsilon} \left( z^{-2} + \varepsilon \right) |\hat{w}_{\boldsymbol{k}} \hat{T}_{\boldsymbol{k}}|\textrm{d}z.
    \label{eq:sk_fs_e}
\end{align}
Then, we estimate the two integrals at the boundaries in \eqref{eq:sk_fs_e}. Starting with the integral at the lower boundary, given that $\varepsilon\leq \frac13$ and $\delta \leq \frac13$, we take $\delta \varepsilon^{1/2} \leq 1/\delta$. Substituting for $\hat{w}_{\boldsymbol{k}}$ by use of \cref{lem:w_vort_est}, for $\hat{T}_{\boldsymbol{k}}$ from \eqref{eq:T_est} and an application of Youngs' inequality gives
\begin{align}
    \left(\frac{1}{\delta} + \frac{\delta \varepsilon^{1/2}}{3} \right) \int^{\delta}_0 |\hat{w}_{\boldsymbol{k}} \hat{T}_{\boldsymbol{k}}|\textrm{d}z  &\leq \frac{10}{9\delta} \int^{\delta}_0 |\hat{w}_{\boldsymbol{k}}| |\hat{T}_{\boldsymbol{k}}|\textrm{d}z \leq \frac{4 c_1 k^{1/2}}{9} \delta^{3/2} \lVert \hat{\zeta}_{\boldsymbol{k}} \rVert_2 \lVert \hat{T}'_{\boldsymbol{k}} \rVert_2 \nonumber \\ 
    &\leq \frac{\beta}{2} \lVert \hat{T}'_{\boldsymbol{k}} \rVert_2^2 + \frac{8 c_1^2 k }{81 \beta}  \delta^3 \lVert \hat{\zeta}_{\boldsymbol{k}} \rVert_2^2. 
    \label{eq:lower_int_fs}
\end{align}
Next, we consider the integral at the upper boundary. Since the integral is over the open interval of $(1-\varepsilon,1)$ we have the estimate that $z^{-2} + \varepsilon \leq (1-\varepsilon)^{-2} + \varepsilon \leq 3$, then the use of  \cref{lem:w_vort_est},  \eqref{eq:w_T_est_u} and Youngs' inequality gives
\begin{align}
    \frac{\delta}{3\, \varepsilon^{1/2}} \int^{1}_{1-\varepsilon} \left( z^{-2} + \varepsilon \right) |\hat{w}_{\boldsymbol{k}} \hat{T}_{\boldsymbol{k}}|\textrm{d}z &\leq \frac{\delta}{ \varepsilon^{1/2}}\int^{1}_{1-\varepsilon} |\hat{w}_{\boldsymbol{k}}||\hat{T}_{\boldsymbol{k}} | \textrm{d}z \leq \frac{2}{5 } c_1 k^{1/2}  \delta \varepsilon^{2} \lVert \hat{\zeta}_{\boldsymbol{k}} \rVert_2 \lVert \hat{T}'_{\boldsymbol{k}} \rVert_2 \nonumber \\
    &\leq \frac{\beta}{2} \lVert \hat{T}'_{\boldsymbol{k}} \rVert_2^2 +  \frac{2c_1^2 k }{25 \beta} \delta^2 \varepsilon^4 \lVert \hat{\zeta}_{\boldsymbol{k}} \rVert_2^2 .
    \label{eq:upper_int_fs}
\end{align}
Given the estimates in \eqref{eq:B_sf_low} and \eqref{eq:2norm_e} we have estimates for $\beta$ from above and below, where $\beta$ is given by \eqref{e:b-choice} of
\begin{equation}
    \label{eq:beta_est_s}
   \frac{2\sqrt{2}}{9}\delta \leq \beta \leq 2 \delta\, . 
\end{equation}
Taking \eqref{eq:lower_int_fs}, \eqref{eq:upper_int_fs} and the lower bound in \eqref{eq:beta_est_s}, the $\mathcal{S}_{\boldsymbol{k}}$ is estimated as
\begin{align}
    \label{eq:sk_fs_simp}
    \mathcal{S}_{\boldsymbol{k}}\{ \hat{w}_{\boldsymbol{k}}, \hat{T}_{\boldsymbol{k}} \} \geq \left( \frac{2 \sqrt{2} \delta k^4}{9 R^2} + \frac{\delta \varepsilon^{1/2}}{3 R} - \frac{2\sqrt{2}c_1^2 k \delta^2}{9 } - \frac{9\sqrt{2}c_1^2 k \delta \varepsilon^4}{50} \right) \lVert \hat{\zeta}_{\boldsymbol{k}} \rVert_2^2   .
\end{align}
The spectral constraint, $\mathcal{S}_{\boldsymbol{k}}\geq0 $, is now guaranteed when \eqref{eq:sk_fs_simp} is non-negative for all wavenumbers. To proceed we make the choice that $\delta = \frac{81}{100}\varepsilon^4$, such that we require
\begin{equation}
\label{eq:cond_fs}
    \frac{  k^4}{ R^2} + \frac{ (360)^{1/4} }{4} \frac{\delta^{1/8} }{ R} - 2 c_1^2 k \delta  \geq 0.
\end{equation}
The condition in \eqref{eq:cond_fs} has one negative term $O(k)$, balanced for large wavenumbers by the $O(k^4)$ term and for small wavenumbers by the $O(1)$ term. Noticing that \eqref{eq:cond_fs} is convex in $k$ and has a minimum in $k$ in terms of the remaining variables. Differentiating the expression on the left-hand side of \eqref{eq:cond_fs}, setting it to zero and rearranging, we find
\begin{equation}
    k_m = \left(\frac{c_1^2}{2}\right)^{1/3} \delta^{1/3} R^{2/3}.
\end{equation}
Substituting for $k_m$ in \eqref{eq:cond_fs} and $c_1$ from \cref{lem:w_vort_est}, and rearranging then
\begin{equation}
    \label{eq:delt_fs}
    \delta \leq \left( \frac{2^{49}5^3 }{3^{30}} \right)^{2/29} R^{-40/29} ,
\end{equation}
which is the necessary condition to ensure that the spectral constraint is satisfied.

\subsection{Bound on \texorpdfstring{$\mathcal{F}_B$}{test 1} }

Taking $\delta$ as large as possible in \eqref{eq:delt_fs} gives the condition with the largest freedom that satisfies the spectral constraint. Finally, substituting back into \eqref{eq:U_fas} while remembering that $\mean{wT} \leq U$ gives
\begin{equation}
    \wT \leq \frac12 - m_1 R^{-40/29}   + m_2
 R^{-35/29} \ln{(1 - m_3 R^{-10/29})} ,
\end{equation}
where $m_1 = \frac{1}{6} \left( \frac{2^{49} 5^3}{3^{30}} \right)^{2/29} = 1.4953$, $m_2 = \frac{2^{3/4}}{3^{5/2}5^{1/4}}(6m_1)^{7/8} = 0.1026 $ and $m_3 = \left(\frac{20}{27} m_1\right)^{1/4} = 0.6555$ .
By use of \eqref{eq:Flux_def} we obtain that
\begin{equation}
    \mathcal{F}_B \geq m_1 R^{-40/29}   - m_2
 R^{-35/29} \ln{(1 - m_3 R^{-10/29})} ,
\end{equation}
the desired result.

\section{Numerical investigation}
\label{sec:num_res}

We present numerical simulations of uniform internally heated convection as defined in \eqref{e:governing-equations} in a two-dimensional system. The numerical results supplement the bounds by providing insight into the mathematical results. Our primary interest is the value of $\mathcal{F}_B$ and $\volav{T}$ after the simulation reaches a statistically stationary state. We use no-slip and stress-free boundary conditions with $R$ ranging from $10^4$ to $10^9$. We solve the system of equations with Dedalus \cite{burns2020}, using pseudospectral spatial derivatives and a second-order semi-implicit BDF time-stepping scheme \cite{wang_variable_2008}. Numerical convergence was verified in the stationary state by ensuring at least ten orders of magnitude between the lowest and highest order in the power spectrum for both dimensions. We use Chebyshev polynomials in the finite vertical dimension and Fourier bases in the periodic horizontal dimension. The aspect ratio of the horizontal to vertical dimensions is 4:1. The results were initially benchmarked against Goluskin\& van der Poel (2016)\cite{Goluskin2016ih} with $Pr=1$ runs. The parameters and data attained are in \cref{tab:sim}.

\begin{figure}
    \centering
    \includegraphics[scale=0.9]{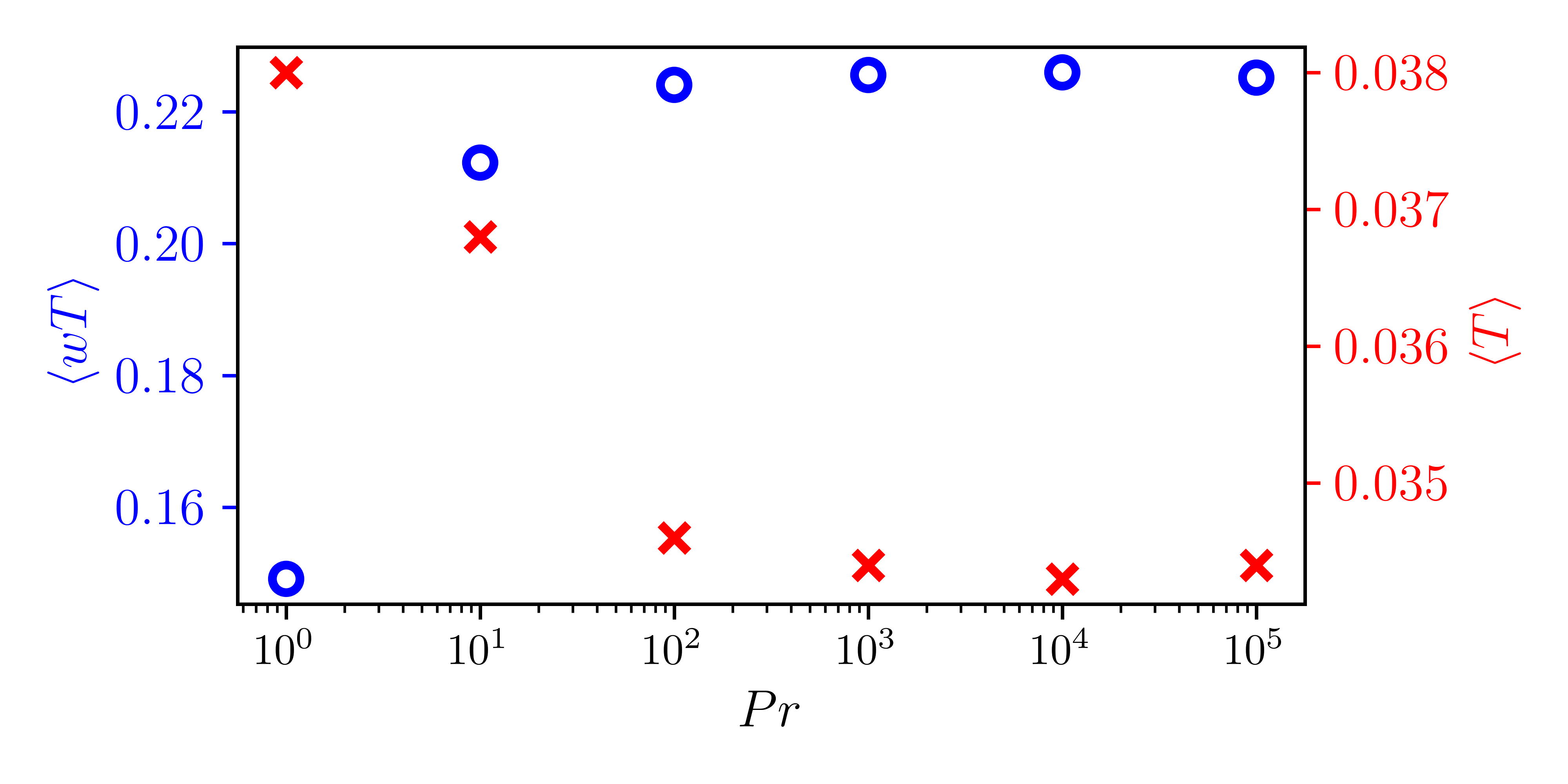}
    \caption{Plots of the mean temperature $\volav{T}$ (right vertical axis, \mycross{colorbar12}) and the mean vertical convective heat transport $\volav{wT}$ (left vertical axis, \mycirc{blue}) as a function of the Prandtl number, $Pr$, for a Rayleigh number of $ R = 2\times10^7$ and no-slip boundary conditions. }
    \label{fig:pr}
\end{figure}

Additionally, we performed simulations for different Prandtl numbers in Figure \ref{fig:pr}, ranging from $Pr=1$ to $Pr=10^5$ for fixed Rayleigh numbers. In which case the momentum equation \eqref{nondim_momentum} is given by
\begin{equation}
    \partial_t \boldsymbol{u} + \boldsymbol{u}\cdot\nabla\boldsymbol{u} + \nabla p = Pr\,\Delta\boldsymbol{u} + Pr\,R\,T\uvec{z}.
\end{equation}
We found no significant statistical difference in $\volav{wT}$ and $\volav{T}$ for Prandtl higher than 100 in the range of $R$ explored. The results of this parameter study motivated the choice of $Pr=10^3$ for the simulations. Similar findings have been observed and implemented in other numerical studies \cite{davies2009convection}.

\begin{figure}
    \centering
    \includegraphics[width=\textwidth]{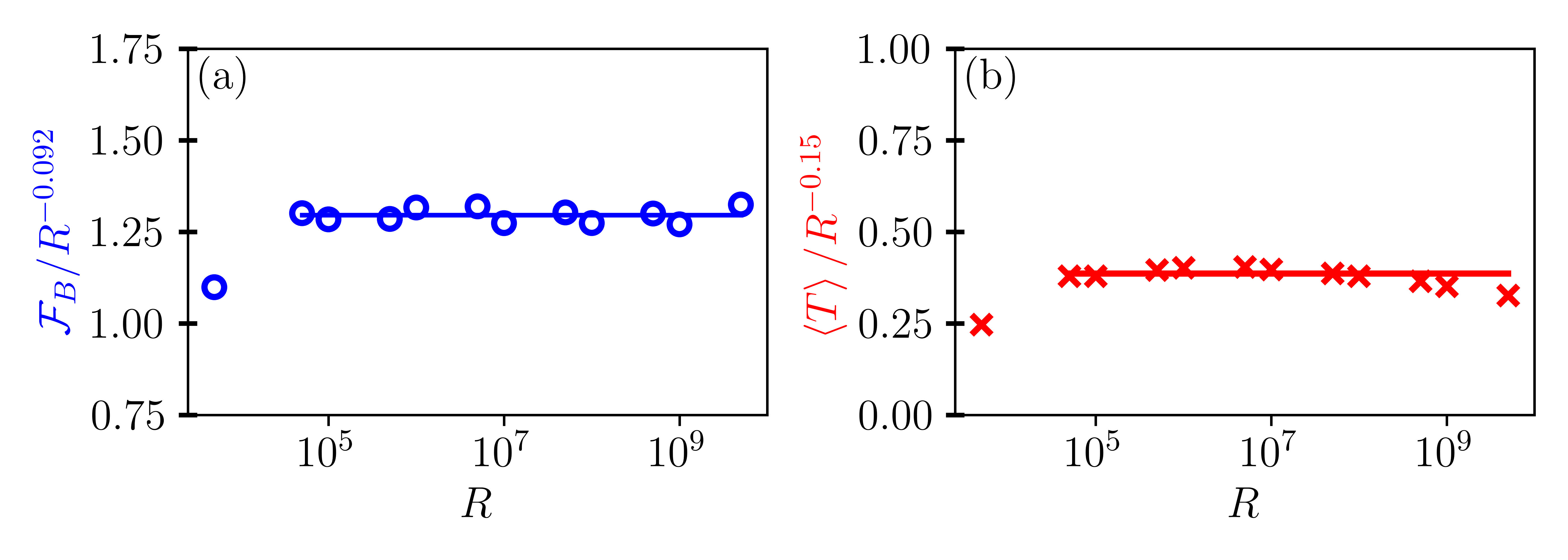}
    \caption{Compensated plots of the horizontally averaged heat flux out of the bottom boundary, $\mathcal{F}_B$, in panel \textrm{(a)}, where the straight line corresponds to $R^{-0.092}$ and the mean temperature $\volav{T}$ in panel \textrm{(b)} where the straight line corresponds to $R^{-0.15}$, as functions of the Rayleigh number, $R$. The simulations are carried out with $Pr=10^3$ for isothermal no-slip boundary conditions, with the data provided in full in \cref{tab:sim}.}
    \label{fig:ns}
\end{figure}
\begin{figure}
    \centering
    \includegraphics[width=\textwidth]{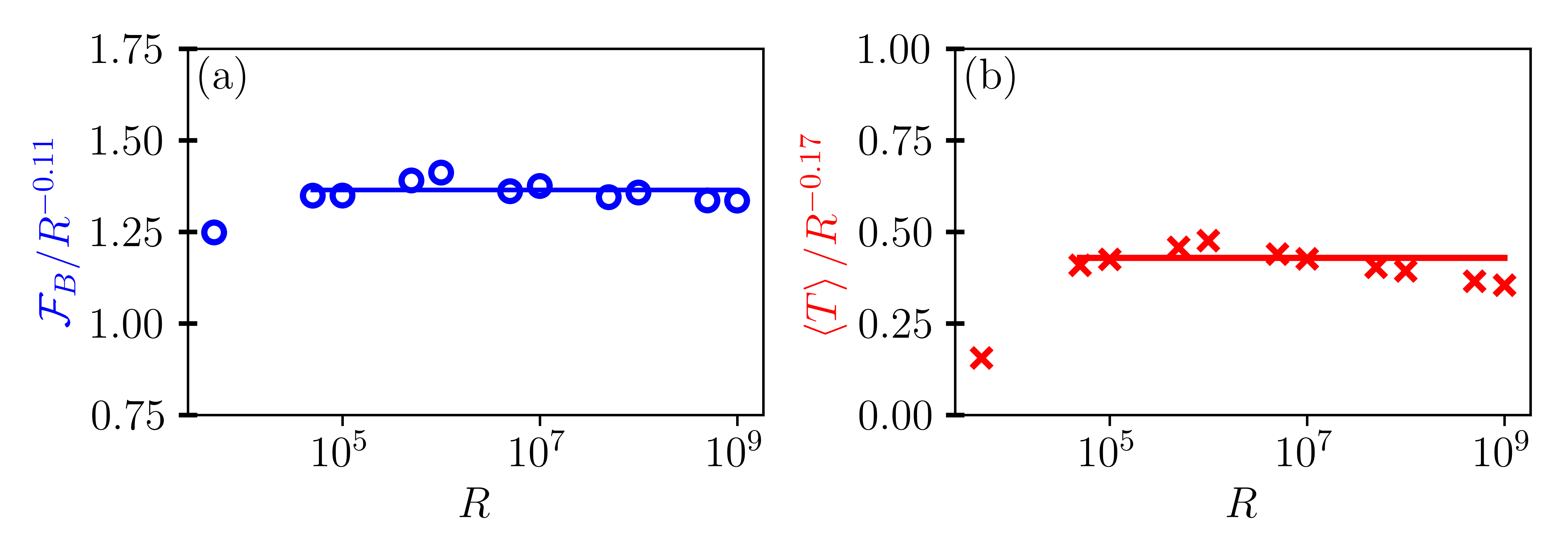}
    \caption{Compensated plots of the horizontally averaged heat flux out of the bottom boundary, $\mathcal{F}_B$, in panel \textrm{(a)}, where the straight line corresponds to $R^{-0.11}$ and the mean temperature, $\volav{T}$, in panel \textrm{(b)}, where the straight line corresponds to $R^{-0.17}$ as a function of the Rayleigh number. The simulations are carried out with $Pr=10^3$ for isothermal stress-free boundary conditions, with the data provided in full in \cref{tab:sim}.}
    \label{fig:fs}
\end{figure}
The main results are in the compensated plots \cref{fig:ns} and \cref{fig:fs}, where we identify from our data a scaling for $\mathcal{F}_B$ and $\volav{T}$ with $R$ in the turbulent regime. The first data points at $R = 5\times10^4$ are ignored when determining the exponent of the Rayleigh scaling. For no-slip boundary conditions we find that, $\mathcal{F}_B \sim R^{-0.092}$ and $\volav{T}\sim R^{-0.15}$. In the stress-free case, the value of the exponents of both quantities is larger at $\sim R^{-0.11}$ and $\sim R^{-0.17}$, respectively.

\begin{figure}
    \centering
    \includegraphics[width=\textwidth]{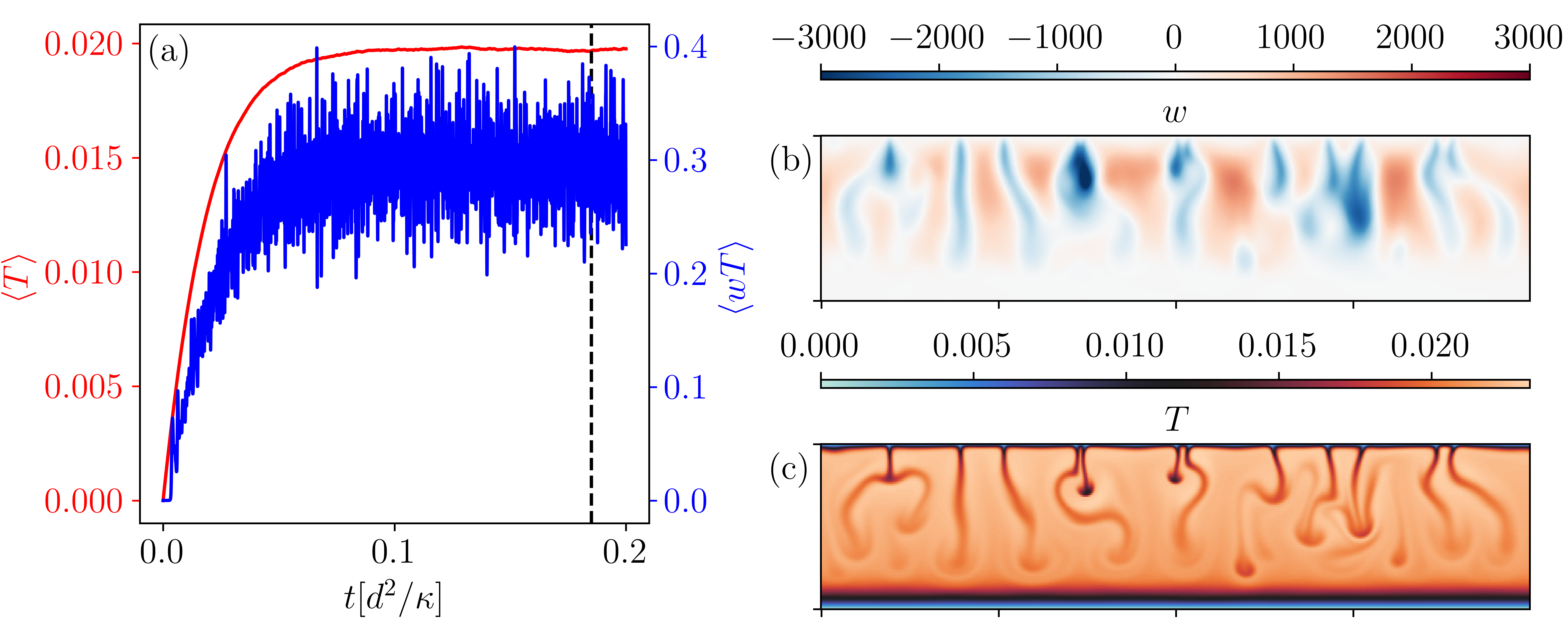}
    \caption{\textrm{(a)} Plot of the evolution with time of the mean vertical convective heat flux $\volav{wT}$ and mean temperature $\volav{T}$. The vertical dashed line (\dashedrule) at $t = 0.18 \; [d^2/\kappa]$ corresponds to the point at which we display the snapshots in \textrm{(b)} and \textrm{(c)}. Panel \textrm{(b)} is a contour plot of the vertical velocity $w$ and panel \textrm{(c)} the temperature field $T$ for $R = 5 \times 10^8$ and $Pr = 10^3$, after statistical convergence of the mean values for stress-free boundary conditions.}
    \label{fig:snap}
\end{figure}
In panel \textrm{(a)}, the evolution of the $\volav{T}$ and $\volav{wT}$ are shown as the simulation progresses.  In \cref{fig:snap}, a snapshot of the vertical velocity field and temperature field are shown in a contour plot for $R=5\times10^8$ in panels \textrm{(b)} and \textrm{(c)}. The contour plots demonstrate that the flow is characterised by downward plumes from the upper unstably stratified thermal boundary layer as per previous studies \cite{worner_direct_1997,goluskin2012convection}.

\section{Discussion}
\label{sec:conc}

We prove new bounds for uniform internally heated convection (IHC) at infinite Prandtl number between isothermal plates with no-slip and stress-free boundary conditions. More precisely, we prove new bounds on the horizontally averaged heat flux out of the boundaries, $\mathcal{F}_B$ and $\mathcal{F}_T$ in terms of the Rayleigh number, $R$, by the background field method formulated in terms of auxiliary functionals. Then, we performed numerical experiments to study $\mathcal{F}_B$ and the mean temperature, $\mean{T}$, for $R$ between $10^4$ and $10^9$ in a two-dimensional domain with Dedalus. In this section, we discuss the significant elements of the results.

For no-slip boundaries we prove that, $\mathcal{F}_B \geq c_1 R^{-2/3} - c_2 R^{-1/2}\ln(1-c_3 R^{-1/3})$, where $c_1 = 0.1555$, $c_2= 0.7104$ and $c_3 = 1.5334$, which improves on the previously known best bound of, $\mathcal{F}_B \gtrsim R^{-2}$ from Arslan et al.(2023)\cite{Arslan2023}. While for stress-free boundary conditions we prove, $\mathcal{F}_B \geq c_1R^{-40/29} - c_2 R^{-35/29} \ln{(1-c_3 R^{-10/29})}$ where $c_1 = 0.0131$, $c_2 = 0.0312$ and $c_3 = 0.6301$. The two features of our proof that lead to the improved bound is (i) the use of pointwise estimates on $w$ and $w''$ from previous works on Rayleigh-B\'enard convection (RBC)\cite{Doering2001,Whitehead2012} (ii) a background temperature field with a $1/z$ behaviour near the upper boundary.

We highlight that while the momentum equation for RBC and IHC are identical, the mean convective heat transport, $\wT$, describes a different physical feature of the turbulence. While $\wT$ in RBC is unbounded and directly determines the Nusselt number, in IHC, it is bounded (by $\frac12$ due to the choice of non-dimensionalisation) and relates to the heat flux out of the domain $\mathcal{F}_B$. In fact, for uniform IHC, $\mathcal{F}_B = \frac12 - \wT$ and a lower bound on $\mathcal{F}_B$ is obtained from an upper bound on $\wT$. In general, it is difficult to prove a lower bound on $\wT$ with quadratic auxiliary functionals. The zero lower bound of $\wT$ can always be saturated by trivial solutions of the system, which would need to be excluded from the set of fields being optimised over. An approach that achieves this feature is the so-called optimal wall-to-wall transport method\cite{Tobasco2017,doering2019optimal} that has recently demonstrated interesting results for RBC\cite{kumar2022b}.

In RBC, Doering \& Constantin (2001) \cite{Doering2001}, by the use of \cref{lem:DC_inf} prove that when $Pr=\infty$, $Nu \lesssim Ra^{2/5}$, with a background temperature field, $\tau(z)$, that is linear in the boundary layers and a constant elsewhere. However, by taking $\tau(z)$ that is logarithmic in the bulk of the domain, i.e. $\sim\ln(z/(1-z))$, the bound is improved to $Nu \lesssim Ra^{1/3}$ up to logarithmic factors, as first demonstrated in Doering, Otto \& Reznikoff (2006)\cite{Doering2006}. In addition to a logarithmic $\tau(z)$, the $Ra^{1/3}$ bound uses an integral inequality of Hardy-Rellich type as opposed to the pointwise estimate in \cref{lem:DC_inf}. Furthermore, logarithmic constructions of $\tau(z)$ for RBC are known to be optimal at $Pr=\infty$ \cite{choffrut2016upper,Nobili2017}. In contrast, for IHC, the previous $O(R^{-2})$ bound in Arslan et al. (2023) \cite{Arslan2023} used logarithmic $\tau(z)$ along with Hardy-Rellich inequalities, while our improvement in this work does not. The optimal $\tau(z)$ for IHC between isothermal boundaries at $Pr=\infty$ is unknown, and the discrepancy in the constructions of $\tau (z)$ between RBC and IHC, and our results, indicate that improvements to the bound on $\mathcal{F}_B$ should be possible. Future work would involve numerically optimising over a finite range of $R$ to determine the optimal $\tau(z)$ and $\lambda(z)$.

The second and novel aspect of the $\tau(z)$ in \eqref{eq:tau_c} and \eqref{eq:tau_fs}, is the $1/z$ behaviour near the upper boundary, which introduces the logarithmic term in the bounds on $\mathcal{F}_B$. Given that the upper boundary layer width $\varepsilon$ is less than 1, it follows that $-\delta \varepsilon^{-1/2} \ln(1-\varepsilon)$, is positive and improves the bound for finite $R$. In the limit of $\varepsilon\rightarrow 0$, the logarithmic term is $O(\delta\varepsilon^{1/2})$ and the bounds are given by the $O(\delta)$ contributions, giving that $\mathcal{F}_B \gtrsim R^{-2/3}$ and $\mathcal{F}_B \gtrsim R^{-40/29}$. The choice of $\tau(z)$ made here will also improve the bounds on $\wT$ at finite $Pr$ since the expression for $U$ in \eqref{eq:U} is identical. A final feature of the background fields is that the best bound is obtained when the boundary layer widths are different. In the case of no-slip boundaries, we set $\delta = \varepsilon^2$ and $\delta = \varepsilon^4$ for stress-free boundaries. The choice for no-slip boundaries matches predictions in heuristic arguments \cite{Arslan2021,Wang2020}, and the difference in boundary layer widths is observed in all numerical studies  (\cref{fig:snap}).

\begin{figure}
    \centering
    \includegraphics[width=\textwidth]{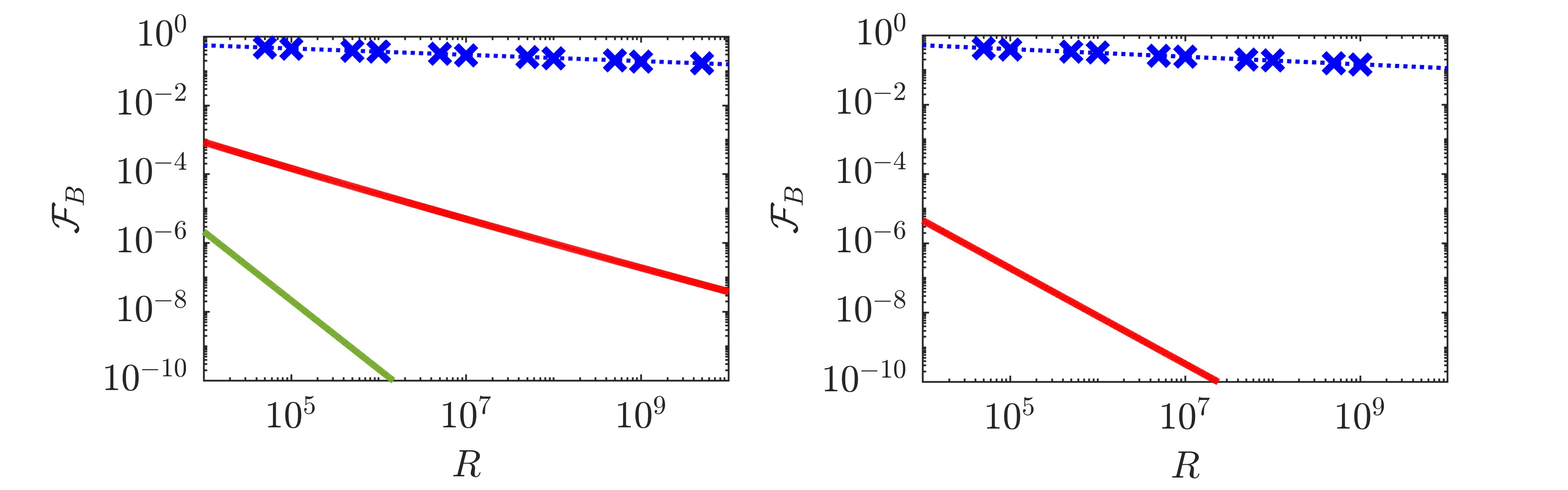}
    \begin{tikzpicture}[overlay]
        \node at (-14,4.6) {\textrm{(a)}};
        \node at (-6.8,4.6) {\textrm{(b)}};
    \end{tikzpicture}
    \caption{Plots of the data from numerical simulations with the rigorous bounds on $\mathcal{F}_B$ against $R$. In panel \textrm{(a)}, the results for no-slip boundaries from \cref{tab:sim} are plotted with blue crosses ({\mycross{myblue}}) and a dotted blue line  ({\color{myblue}\dottedrule}) shows the slope $R^{-0.092}$, the red solid line ({\color{colorbar12}\solidrule}) shows the lower bound in \eqref{thm:ns} and the green solid line  ({\color{matlabgreen}\solidrule}) the previously known best bound from Arslan et al. (2023)\cite{Arslan2023}. In panel \textrm{(b)}, the results for stress-free boundaries from \cref{tab:sim} are plotted in blue crosses ({\mycross{myblue}}) with a dotted blue line ({\color{myblue}\dottedrule}) showing the slope $R^{-0.11}$ and the red solid line ({\color{colorbar12}\solidrule}) shows the lower bound in \eqref{thm:sf}.  }
    \label{fig:bounds_c}
\end{figure}
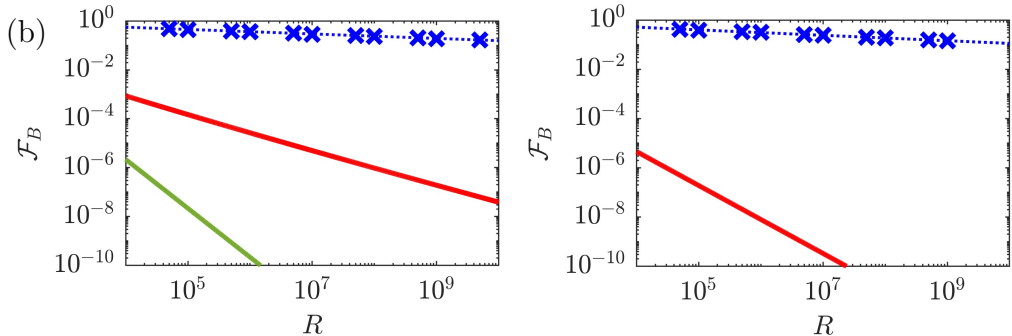

\begin{figure}
    \centering
    \includegraphics[scale=0.9]{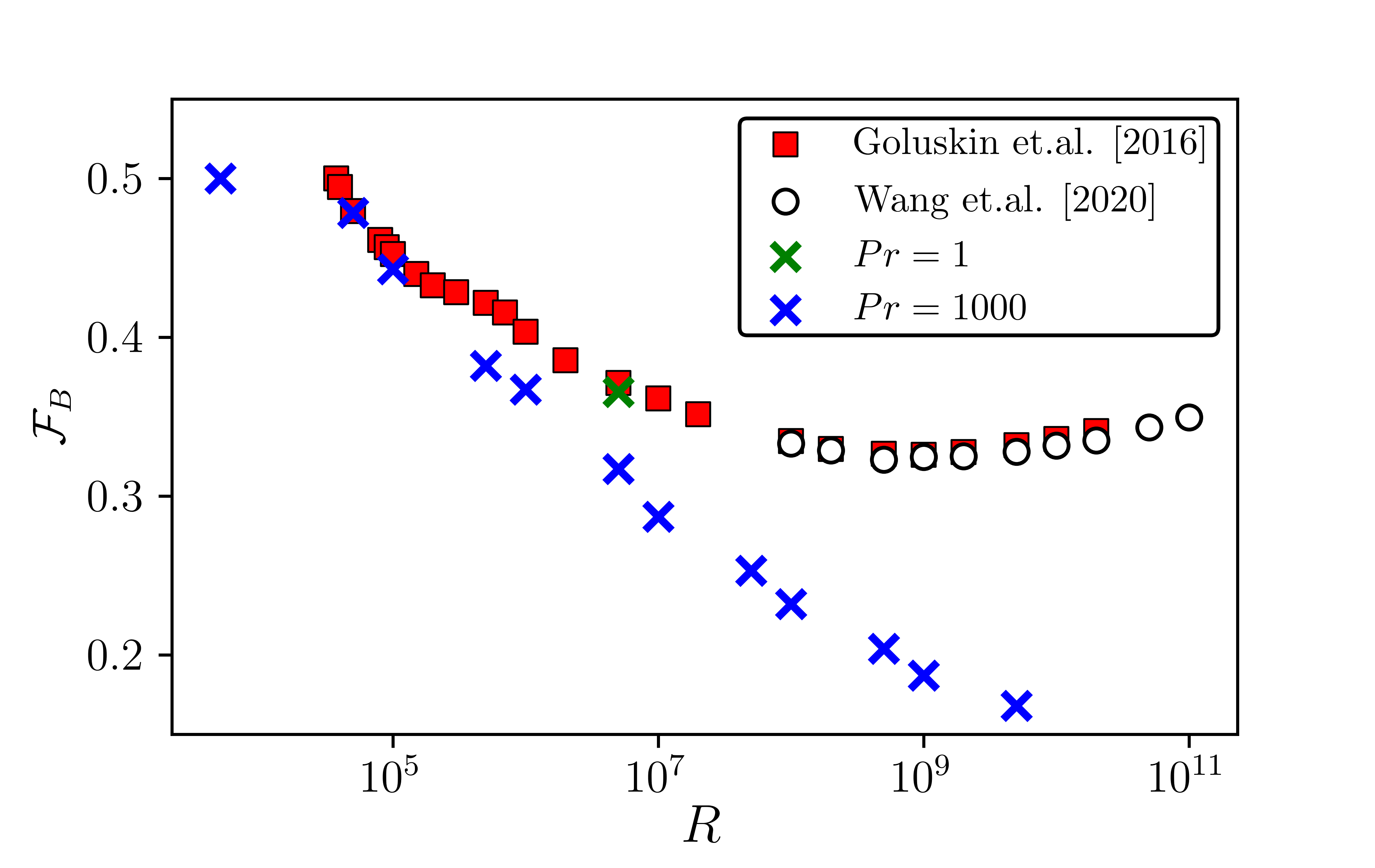}
    \caption{Plot of the heat flux out of the bottom boundary, $\mathcal{F}_B$, for different Rayleigh numbers, $R$, between isothermal no-slip boundaries in two dimensions. The green cross ({\mycross{matlabgreen}}) represents data at $Pr = 1$ and the blue cross ({\mycross{myblue}}) the data at $Pr = 10^3$ from \cref{tab:sim}. Results of Goluskin \& van der Poel (2016)\cite{Goluskin2016ih} are shown in red squares (\mysquare{colorbar12}) and Wang et al. (2020) \cite{Wang2020} in white circles (\mycirc{white}) at $Pr = 1$ for the same boundary conditions in two-dimensions.}
    \label{fig:comp_res}
\end{figure}

The main results of this work are the new rigorous bounds on $\mathcal{F}_B$, but we carried out a numerical study, which we now discuss, to gauge how conservative our results are. In the results (\cref{fig:ns}\textrm{(a)}), we see that $\mathcal{F}_B \sim R^{-0.092}$, and the exponent of $R$ is a factor of $7$ lower than the rigorous bound in \eqref{thm:ns}. Both are plotted in \cref{fig:bounds_c}\textrm{(a)} with blue crosses ({\mycross{matlabblue}}) and a red line ({\color{matlabred}\solidrule}) respectively. For comparison, Goluskin \& van der Poel (2016)\cite{Goluskin2016ih} report that $\mathcal{F}_B \sim R^{-0.055}$ at $Pr=1$, where their exponent is $39.6\%$ lower than our value in \cref{fig:ns}. The authors also find larger values of $\mathcal{F}_B$ and a non-linear behaviour for two, compared to three dimensions for their runs above $R\sim10^{9}$. Wang et al. (2020)\cite{Wang2020} report a numerical study in two dimensions but do not provide a scaling for $\mathcal{F}_B$ due to the small range of $R$ covered. \cref{fig:comp_res}  plots both of the aforementioned works with our simulations, with a $Pr=1$ run in a green cross ({\mycross{matlabgreen}}) for comparison. Unlike results for $Pr=1$, in \cref{fig:comp_res} we do not observe an increase in $\mathcal{F}_B$ in the $Pr=10^3$ runs ({\mycross{myblue}}) when $R$ gets large. This is an interesting difference between the turbulence at high $R$ for IHC in 2D and is left to future Prandtl number studies.
A possible limitation in obtaining sharp bounds for $\mathcal{F}_B$ could be the choice of auxiliary functional. The use of higher-than-quadratic functionals is analytically intractable such that using additional constraints beyond the minimum principle on $T$ could be the feasible route to improvement. Alternatively, an optimal $\tau(z)$ in the proof might give a sharp bound, however, this was not the case or IHC at finite $Pr$\cite{Arslan2021,Arslan2021a}.

To our knowledge, no further numerical studies for IHC report results on $\mathcal{F}_B$, instead focusing on $\mean{T}$, taken as a proxy for the Nusselt number.  In \cref{fig:ns}\textrm{(b)} and \cref{fig:fs}\textrm{(b)} we find that the mean temperature scales as $R^{-0.15}$ and $R^{-0.17}$, for no-slip and stress-free boundaries. The results for $\mean{T}$ in \cref{fig:fs} can be compared to Sotin \& Labrosse (1999) \cite{sotin1999}, where the authors carry out 3D simulations at infinite $Pr$ and find that $\mean{T} \sim  R^{-0.234}$ for stress-free boundaries. Their results have an exponent $80\%$ larger than the value we find. The authors consider a model with uniform internal and boundary heating, which could account for the difference, aside from the dimension of the simulation.
The main results of the stress-free simulations are $\mathcal{F}_B \sim R^{-0.11}$, which, as compared in \cref{fig:bounds_c}\textrm{(b)}, has an exponent a factor of $20$ smaller than our rigorous bound. For stress-free plates, vorticity is not produced at the boundaries, and no vortex stretching occurs in the bulk of the fluid. Therefore, the efficiency of convective heat transport should increase and does so when comparing our results between the two kinematic boundary conditions.

A feature of the background profiles in our proof is the role of $A$ defined in \eqref{e:A-choice} and \eqref{eq:A_2}, which gives a bound on $\wT$ that asymptotes to $\frac12$ from below and defines the balance parameter $\beta$ (a $R$ dependent parameter). The dependence of $\beta$ on $R$ is a critical element of our bound on $\wT$ and the main variation with RBC, where $\beta$ is a constant.
The magnitude of $A$ is crucial to a bound for stress-free boundaries, where $A$ is the value of $\tau'$ in the bulk and necessary in satisfying the spectral constraint. Additionally, when satisfying the spectral constraint in the stress-free case, we minimise over $k$ and find that $k_m$ is $\sim R^{18/87}$. For $\mean{T}$ in IHC, Whitehead \& Doering (2012)\cite{Whitehead2012}, for whom the proof follows with the same approach, find that $k_m \sim R^{3/17}$. The discrepancy between $k_m$ could indicate that further improvements are possible to the result proven in the stress-free case.

Finally, the problem considered here is that of uniform IHC, although, for geophysical applications like Mantle convection, the internal heating is non-uniform. Therefore, the effects of spatially varying heat sources on heat transport are of interest to applications, and a natural question would be on the nature of rigorous bounds for the emergent quantities that depend on the heating location in addition to $R$ or $Pr$.
Recent mathematical, numerical and experimental work has looked at the case of net zero internal heating and cooling \cite{Lepot2018,Song2022,bouillaut2022,kazemi2022}, to explore the different regimes of turbulence. However, if the fluid is heated and cooled, the minimum principle no longer holds and the physics, and consequently quantities of interest, change. Nevertheless, the emergent quantities can be studied if the arbitrary heating is strictly positive \cite{Arslan2024}, wherein a link is possible between the rigorous bounds of IHC and that of previous studies of RBC.

\begin{acknowledgments}
We acknowledge T. Sternberg, A. Wynn, J. Craske and G. Fantuzzi for their helpful discussions. The authors acknowledge funding from the European Research Council (agreement no. 833848-UEMHP) under the Horizon 2020 program. A.A. acknowledges the Swiss National Science Foundation (grant number 219247) under the MINT 2023 call. 
\end{acknowledgments}


\appendix

\section{Data of the numerical results}

In this section, we present the exact data for the simulations carried out as described in \cref{sec:num_res}.
\begin{table*}
\caption{\label{tab:sim}%
Details of the 2D numerical simulations carried out with Dedalus. Each row corresponds to an individual run and the columns from left to right show the Rayleigh number, the Prandtl number, the number of terms in the vertical Chebyshev and horizontal Fourier series, the simulations time, the volume-averaged temperature, the volume-averaged product of the vertical velocity and temperature and the kinematic boundary conditions.
}
\begin{ruledtabular}
\begin{tabular}{lcccccc}
$R$ & $Pr$ & $N_x$, $N_z$ & Time[$d^2/\kappa$] & $\left<T\right>$ & $\left<wT\right>$ & BC \\
\midrule
\midrule
$5 \times 10^3$ & 1000 & 256,64 & 10 & 0.0718 & 0 & no-slip \\
$5 \times 10^4$ & 1000 & 256,64 & 10 & 0.0790 & 0.0782 & no-slip \\
$1 \times 10^5$ & 1000 & 256,64 & 10 & 0.0714 & 0.0571 & no-slip \\
$5 \times 10^5$ & 1000 & 256,64 & 8 & 0.0590 & 0.162 & no-slip \\
$1 \times 10^6$ & 1000 & 256,128 & 7 & 0.0542 & 0.133 & no-slip \\
$5 \times 10^6$ & 1000 & 256,128 & 1 & 0.0432 & 0.183 & no-slip \\
$1 \times 10^7$ & 1000 & 512,128 & 2 & 0.0384 & 0.213 & no-slip \\
$5 \times 10^7$ & 1000 & 512,128  & 1 & 0.0296 & 0.242 & no-slip \\
$1 \times 10^8$ & 1000 & 512,128 & 1 & 0.0261 & 0.268 & no-slip \\
$5 \times 10^8$ & 1000 & 512,128 & 0.5 & 0.0198 & 0.275 & no-slip \\
$1 \times 10^9$ & 1000 & 512,256 & 0.5 & 0.0174 & 0.313 & no-slip \\
$5 \times 10^9$ & 1000 & 512,256 & 0.05 & 0.0128 & 0.329 & no-slip \\
$5 \times 10^3$ & 1000 & 256,64 & 10 & 0.0373 & 0 & stress-free \\
$5 \times 10^4$ & 1000 & 256,64 & 5 & 0.0664 & 0.0782 & stress-free \\
$1 \times 10^5$ & 1000 & 256,64 & 5 & 0.0614 & 0.108 & stress-free \\
$5 \times 10^5$ & 1000 & 256,128 & 2 & 0.0504 & 0.161 & stress-free \\
$1 \times 10^6$ & 1000 & 512,128 & 2 & 0.0467 & 0.180 & stress-free \\
$5 \times 10^6$ & 1000 & 512,128 & 1 & 0.0329 & 0.241 & stress-free \\
$1 \times 10^7$ & 1000 & 512,128 & 1 & 0.0284 & 0.257 & stress-free \\
$5 \times 10^7$ & 1000 & 512,256 & 0.5 & 0.0205 & 0.299 & stress-free \\
$1 \times 10^8$ & 1000 & 512,256 & 0.5 & 0.0178 & 0.312 & stress-free \\
$5 \times 10^8$ & 1000 & 1024,256 & 0.1 & 0.0126 & 0.345 & stress-free \\
$1 \times 10^9$ & 1000 & 1024,256 & 0.05 & 0.0109 & 0.356 & stress-free \\
$2 \times 10^7$ & 1 & 256,64 & 2 & 0.0380 & 0.1491 & no-slip \\
$2 \times 10^7$ & 10 & 256,64 & 2 & 0.0368 & 0.2123 & no-slip \\
$2 \times 10^7$ & 100 & 256,64 & 2 & 0.0346 & 0.2241 & no-slip \\
$2 \times 10^7$ & 1000 & 256,64 & 2 & 0.0344 & 0.2256 & no-slip \\
$2 \times 10^7$ & 10000 & 256,64 & 2 & 0.0343 & 0.2260 & no-slip \\
$2 \times 10^7$ & 100000 & 256,64 & 2 & 0.0344 & 0.2252 & no-slip \\
\end{tabular}
\end{ruledtabular}
\end{table*}

\newpage


\bibliography{bibs.bib}

\end{document}